\documentclass[fleqn,usenatbib]{mnras}
\usepackage[T1]{fontenc}
\usepackage{ae,aecompl}
\usepackage{amsmath}
\usepackage{amssymb}

\usepackage{graphicx}
\graphicspath{{figure/}}

\usepackage{url}
\newcommand{\kms}{\,km\,s$^{-1}$}
\newcommand{\Msun}{\,M$_\odot$}
\newcommand{\sqarcmin}{\,arcmin$^{2}$}
\newcommand{\sqdeg}{\,deg$^{2}$}

\newcommand{\molh}{$\text{H}_2$}

\defcitealias{Yung2019}{Paper~I}
\defcitealias{Yung2019a}{Paper~II}
\defcitealias{Yung2020}{Paper~III}
\defcitealias{Yung2020a}{Paper~IV}
\defcitealias{Yung2021}{Paper~V}

\defcitealias{Bruzual2003}{BC03}
\defcitealias{Gnedin2011}{GK}
\defcitealias{Bigiel2008}{Big}
\defcitealias{Somerville2015}{SPT15}
\defcitealias{Somerville2021}{S21}

\title[SAM forecasts -- VI. Simulated JWST lightcones]{Semi-analytic forecasts for \textit{JWST} -- VI. Simulated lightcones and galaxy clustering predictions}

\author[L. Y. A. Yung et al.]{L. Y. Aaron\ Yung,$^{1}$\thanks{E-mail: aaron.yung@nasa.gov}\thanks{NPP Fellow}
Rachel S.\ Somerville,$^{2}$ Henry C.\ Ferguson,$^{3}$ 
\newauthor Steven L.\ Finkelstein,$^{4}$ Jonathan P.\ Gardner,$^{1}$ Romeel\ Dav\'e,$^{5,6,7}$ Micaela B.\ Bagley,$^{4}$
\newauthor  Gerg\"o\ Popping$^{8}$ and Peter\ Behroozi$^{9}$
\\
$^{1}$Astrophysics Science Division, NASA Goddard Space Flight Center, 8800 Greenbelt Rd, Greenbelt, MD 20771, USA\\
$^{2}$Center for Computational Astrophysics, Flatiron Institute, 162 5th Ave, New York, NY 10010, USA\\
$^{3}$Space Telescope Science Institute, 3700 San Martin Drive, Baltimore, MD 21218, USA\\
$^{4}$Department of Astronomy, The University of Texas at Austin, Austin, TX 78712, USA\\
$^{5}$Institute for Astronomy, University of Edinburgh, Edinburgh EH9 3HJ, UK\\
$^{6}$Department of Physics and Astronomy, University of the Western Cape, Cape Town 7535, South Africa\\
$^{7}$South African Astronomical Observatory, Cape Town 7925, South Africa\\
$^{8}$European Southern Observatory, Karl-Schwarzschild-Strasse 2, D-85748 Garching, Germany\\
$^{9}$Department of Astronomy, University of Arizona, 933 N Cherry Ave, Tucson, AZ 85721, USA
}

\date{Accepted XXX. Received YYY; in original form ZZZ}

\pubyear{2022}

\begin{document}
\label{firstpage}
\pagerange{\pageref{firstpage}--\pageref{lastpage}}
\maketitle

\begin{abstract}
In anticipation of the new era of high-redshift exploration marked by the commissioning of the \emph{James Webb Space Telescope} (\emph{JWST}), we present two sets of galaxy catalogues that are designed to aid the planning and interpretation of observing programs.
We provide a set of 40 wide-field lightcones with footprints spanning approximately $\sim 1,000$\,\sqarcmin\, containing galaxies up to $z = 10$, and a new set of 8 ultra-deep lightcones with 132\,\sqarcmin\ footprints, containing galaxies up to $z\sim12$ down to the magnitudes expected to be reached in the deepest \emph{JWST} surveys.
These mock lightcones are extracted from dissipationless N-body simulations and populated with galaxies using the well-established, computationally efficient Santa Cruz semi-analytic model for galaxy formation. 
We provide a wide range of predicted physical properties, and simulated photometry from NIRCam and many other instruments. 
We explore the predicted counts and luminosity functions and angular two-point correlation functions for galaxies in these simulated lightcones.
We also explore the predicted field-to-field variance using multiple lightcone realizations.
We find that these lightcones reproduce the available measurements of observed clustering from $0.2 \lesssim z \lesssim 7.5$ very well. We provide predictions for galaxy clustering at high redshift that may be obtained from future \emph{JWST} observations. 
All of the lightcones presented here are made available through a web-based, interactive data release portal \url{https://flathub.flatironinstitute.org/group/sam-forecasts}.
\end{abstract}


\section{Introduction}

The launch, deployment, and commissioning of the \emph{James Webb Space Telescope} \citep[\emph{JWST};][]{Gardner2006} marks the beginning of a new era of exploration of the physical origin of our Universe. Its onboard Near-Infrared Camera (NIRCam) possesses unprecedented infrared sensitivity and spatial resolution, which will enable it to image distant, faint objects with great efficiency. It is expected to detect distant galaxies with luminosities  many magnitudes fainter than those discovered by its predecessors the \emph{Hubble} and \emph{Spitzer Space Telescopes}. The Mid-Infrared Instrument (MIRI) has both imaging and spectroscopic capabilities at longer wavelength mid-IR surpassing \emph{Spitzer} and \emph{Herschel} in both spatial resolution and spectral resolution.  The Near-Infrared Imager and Slitless Spectrograph (NIRISS) is capable of effectively obtaining low-resolution spectra of multiple objects at a time, while the Near-Infrared Spectrograph (NIRSpec) will enable high spectral resolution diagnostics with greater sensitivity. These instruments are able to work in parallel to maximize the science yield.

The combination of the large primary mirror and unprecedented infrared sensitivity of \emph{JWST's} scientific instruments will enable the detection of galaxies forming during or prior to the Epoch of Reionization (EoR), providing important information that will help complete our picture of the origins of galaxies and the sources of reionization.
The detection of galaxies of faint apparent magnitude across all epochs of cosmic history will deliver new constraints on the evolution of galaxy physical properties and number densities across cosmic time, as well as insights on the physical processes that shaped these objects.
We also anticipate that \emph{JWST} will, for the first time in history, definitively detect and deliver direct constraints on populations of galaxies forming at $z > 10$, which will provide new tests of galaxy formation models and theories.

Although \emph{JWST} was not designed to survey large areas to deep levels, with the anticipated potential long lifetime that the facility may enjoy, we also anticipate that \emph{JWST} will be able to make important contributions to measuring the spatial clustering of these faint, high redshift galaxies. This can provide an important counterpart to the wider, shallower clustering measurements that will be enabled by ground based facilities and the Euclid and Nancy Grace Roman space telescopes. Galaxy clustering can provide important constraints on the distribution of the underlying dark matter density field, as well as on how galaxies of different types populate halos of different mass (halo occupation).

In order to leverage the rich data accumulated from past observations, \emph{JWST} is expected to revisit the five relatively well-surveyed legacy fields from the Cosmic Assembly Near-infrared Deep Extragalactic Legacy Survey \citep[CANDELS;][]{Grogin2011, Koekemoer2011}.
These legacy fields include the Extended Groth Strip (EGS) from the All-wavelength Extended Groth strip International Survey \citep[AEGIS;][]{Davis2007}, the north and south fields of the Great Observatories Origins Deep Survey \citep[GOODS-N and GOODS-S;][]{Giavalisco2004a}, and the Cosmic Evolution Survey \citep[COSMOS;][]{Scoville2007}, and the UKIDSS Ultra Deep Survey \citep[UDS;][]{Cirasuolo2007}.
In addition, there is also the \emph{Hubble} Ultra Deep Field \citep[HUDF;][]{Ellis2013, Koekemoer2013} and the eXtreme Deep Field \citep[XDF;][]{Illingworth2013}, located within GOODS-S.

Within the first year of operation, \emph{JWST} will execute a number of Early Release Science and Guaranteed Time (GTO) deep extragalactic survey programs.
The Cosmic Evolution Early Release Science (CEERS\footnote{\url{https://ceers.github.io}}) Survey will cover $\sim100$ square arcminutes within the EGS field \citep{Finkelstein2017} with \emph{JWST} imaging and spectroscopy using NIRCam, MIRI, and NIRSpec.
The joint NIRCam--NIRSpec GTO \emph{JWST} Advanced Deep Extragalactic Survey \citep[JADES;][]{Williams2018} will observe the GOODS-S and HUDF field.
In addition, the Cycle 1 General Observer (GO) program Next Generation Deep Extragalactic Exploratory Public (NGDEEP) will dedicate 120 hours to observing the HUDF with NIRISS and imaging the HUDF Parallel 2 field with NIRCam \citep{Finkelstein2021a}. The Public Release IMaging for Extragalactic Research \citep[PRIMER;][]{Dunlop2021} and COSMOS--Web \citep{Kartaltepe2021} will cover over 390 square arcminutes (over COSMOS and UDS) and 2100 square arcminutes (over COSMOS), respectively.
We refer the reader to \citet{Robertson2021} for a comprehensive review of \emph{JWST} approved galaxy evolution programs and the science that is expected to be enabled by them.

Given \emph{JWST}'s limited lifespan and high over-subscription rates for time allocation, optimization of observational strategies including exposure time, pointings, survey strategy, as well as parallel observing, are crucial to maximize the science return.
Detailed mock catalogues and synthetic data can help support the planning and optimization of these observational strategies.

Lightcones are a form of mock galaxy catalogues that mimic the way we observe the real Universe, providing galaxy locations on the plane of the sky (right ascension and declination) as well as redshifts along the line of sight (including both cosmological redshifting and peculiar velocities) along a past lightcone. They more realistically include the continuous sampling of galaxies at different redshifts along the line of sight in a real survey, along with the accompanying impact on the $k$-corrections of the SED, etc, than using a single redshift slice from a box, as is commonly done for simulations. The output from lightcones is also useful as an input to create synthetic images or spectroscopy. It is straightforward to include observational effects such as photometric noise, a point spread function, and observationally based selection criteria \citep{Torrey2015, Rodriguez-Gomez2019, Snyder2019, Marshall2022}.
Thus simulated lightcones are an important tool for both \textit{forecasting} and interpreting high-redshift galaxy surveys. 
In the past, mock lightcones have been created to aid the theoretical interpretation of the CANDELS surveys \citep{Behroozi2019, Somerville2021}.
The lightcones presented in this work have been used in the interpretation of galaxies observed at high redshifts \citep{Finkelstein2021, Tacchella2021} with \emph{Hubble} and \emph{Spitzer}, as well as in the planning for upcoming CEERS and NGDEEP surveys.
Other \emph{JWST} mock catalogues have also been made available, such as those based on \textsc{UniverseMachine} \citep{Behroozi2019, Behroozi2020}, the \emph{JWST} Extragalactic Mock Catalogue \citep{Williams2018}, and the Deep Realistic Extragalactic Model (DREaM) \citep{Drakos2021}.

There are several different methods used in the literature for creating mock catalogues and lightcones. In purely empirical methods such as the JAGUAR models used to create mock catalogues in support of the JADES survey \citep{Williams2018}, observed galaxy properties are interpolated or extrapolated, and there is no underlying physics model nor setting within a $\Lambda$CDM context. In what are sometimes called `semi-empirical' methods (also called sub-halo abundance matching (SHAM) or halo occupation distribution (HOD) models; see \citealp{Wechsler2018}), galaxy properties are mapped onto the properties of dark matter halos such that a set of observational quantities is reproduced \citep{Behroozi2010, Behroozi2019, Moster2013, Moster2018, Wechsler2022}. Both of these methods have the advantage that they are computationally efficient, are not dependent on a specific model for galaxy formation, and are guaranteed to match the observations that were used to calibrate them. However, they have the disadvantage that using them for \emph{forecasts} for new observations is highly uncertain, and they are of limited use for interpretation. Semi-empirical models are typically calibrated using derived physical properties such as stellar masses and star formation rates, which are highly uncertain at high redshifts, leading to models that are nominally calibrated on the same observations, but which have very different predictions for the link between galaxy and dark matter halo properties \citep[see e.g.][]{Yung2019a}. 
We note that this is only a general overview for the semi-empirical modelling approach. These models are designed with different purposes in mind and adopt different calibration criteria. For example, \textsc{UniverseMachine} used high-redshift UV luminosity functions to calibrate galaxy growth at the highest redshifts \citep{Behroozi2019, Behroozi2020} and the semi-empirical model presented in \citet{Behroozi2015} allows forecasts under the assumption that galaxy-halo growth relationships are given by power laws.

Lightcones can also be extracted from numerical hydrodynamic simulations \citep[e.g.][]{Snyder2017}, but due to the computational expense of the underlying simulations, these tend to have limited volume and are able to provide only a small number of realizations. Moreover, all cosmological hydrodynamic simulations must adopt uncertain `sub-grid' recipes to describe processes that cannot be explicitly resolved, such as star formation, stellar feedback, and black hole seeding, accretion, and feedback \citep{Somerville2015a}.

In this work we adopt the well-established `middle way' for modelling galaxy formation using semi-analytic models. Semi-analytic models are built on the backbone of merger trees based on the $\Lambda$CDM structure formation paradigm, either extracted from dissipationless N-body simulations or created using Monte Carlo methods. They include simplified but physically motivated treatments of the main processes that are thought to shape galaxy evolution, namely gas accretion from the intergalactic medium (IGM) into the circumgalactic medium (CGM), gas cooling from the CGM into the interstellar medium (ISM), formation of stars from dense gas in the ISM, stellar feedback and chemical enrichment from massive stars and supernovae, and black hole growth and feedback. These recipes contain free parameters, that are calibrated to match a subset of observations at $z=0$. Although SAMs can be run in a fraction of the computational cost of numerical hydrodynamic simulations, their predictions for many observable galaxy properties have been shown to be in excellent agreement with those of the numerical simulations \citep{Somerville2015a, Pandya2020, Gabrielpillai2021}.
In particular, the predictions of the Santa Cruz SAM used here have been shown to be in excellent agreement with the predicted rest-UV luminosity functions, stellar mass functions and SFR functions from numerical hydrodynamic simulations over the redshift range $6 \lesssim z \lesssim 10$ \citep{Yung2019a,Lovell2020,Vogelsberger2020a}.

This is the concluding paper in the \emph{Semi-analytic forecasts for JWST} series. This series is a collection of papers that provides in-depth, comprehensive predictions for the properties and demographics of galaxies and AGN forming in the ultrahigh-redshift Universe. 
The modelling pipeline used for the series is based on the versatile Santa Cruz semi-analytic models, with additional model components inserted to expand the model's capabilities in making predictions during the EoR. 
In \citet[hereafter \citetalias{Yung2019} and \citetalias{Yung2019a}]{Yung2019, Yung2019a}, we presented predicted photometric and physical properties of high-redshift galaxy populations, which are compared extensively against existing observations and other simulations. 
In \citet[][hereafter \citetalias{Yung2020}]{Yung2020}, we calculated the ionizing photon production rate based on the star formation and chemical enrichment history of individual galaxies, using both single star and binary stellar population models. In \citet[][hereafter \citetalias{Yung2020a}]{Yung2020a}, we combined the SAM-based source model with an analytic reionization model to compute the cosmic reionization history of intergalactic hydrogen and compare with available observational constraints. 
In \citet[][hereafter \citetalias{Yung2021}]{Yung2021}, we computed the contribution of AGN to hydrogen and helium reionization.
In this final paper of the series (Paper VI), we present mock lightcones with area and resolution chosen to be similar to the footprint and depth of upcoming \emph{JWST} ERS, GTO, and Cycle 1 observing programs, which should be of general utility for planning and interpretation of future \emph{JWST} programs as well. In addition to galaxy physical properties, we include an extensive suite of photometry for \emph{JWST} instruments as well as legacy \emph{HST}, \emph{Spitzer}, and ground based filters. These include lightcones with the same mass resolution and characteristics as the mock CANDELS fields presented in \citet{Somerville2021}, as well as a set of new ultra-deep lightcones with smaller footprints, comparable to deeper, smaller area surveys such as NGDEEP. This work is complementary to the companion paper \emph{Semi-analytic forecasts for Roman} paper \citep{Yung2022a}, which presents a suite of 2-deg$^2$ lightcones with depths comparable to the wide-field lightcones from this work.

We show the predicted luminosity functions and counts from these lightcones in comparison with the predictions from our previous work. We also show a significant new science result enabled by the lightcones. We compute projected two point correlation functions in broad redshift bins, and compare with observational measurements from $0.2 \lesssim z \lesssim 7.5$. Encouraged by the excellent agreement that we find, we present predictions for clustering measurements that may be obtained with future \emph{JWST} observations.

All mock catalogues presented in this work series are accessible through the data release portal \url{https://www.simonsfoundation.org/semi-analytic-forecasts-for-jwst/} and the Flatiron Institute Data Exploration and Comparison Hub (Flathub; \url{http://flathub.flatironinstitute.org/SAM_Forecasts/}).

The structure of this paper is summarized as follows:
the galaxy formation model and lightcone construction pipeline are summarized briefly in Section \ref{sec:models}. 
An overview and description of the simulated lightcones are presented in Section \ref{sec:data_release}.
We present the clustering statistics and other main scientific results in Section \ref{sec:results}.
We discuss our findings in Section \ref{sec:discussion}, and a summary and conclusions follow in Section \ref{sec:snc}.

\begin{figure*}
	\includegraphics[width=2.0\columnwidth]{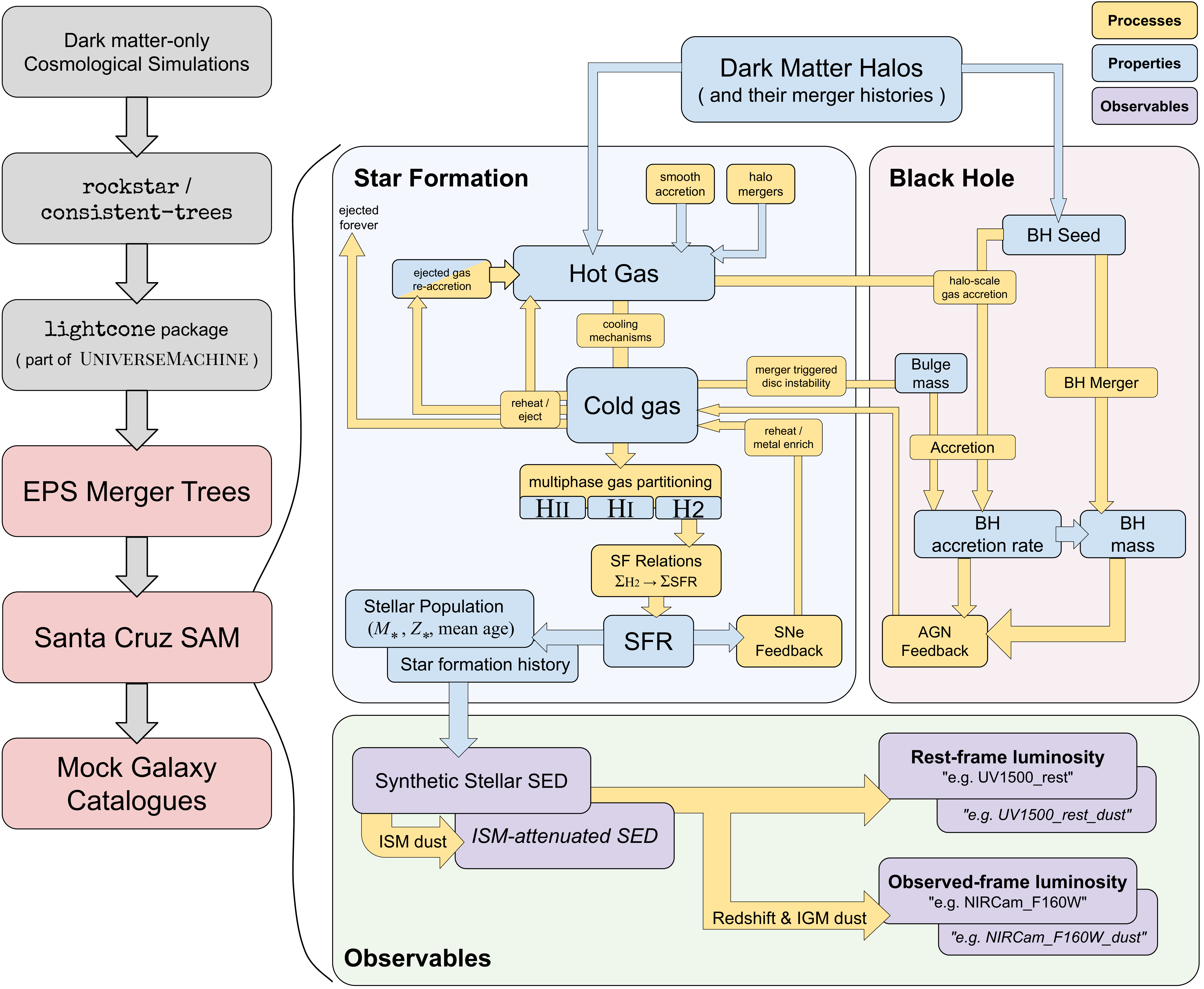}
	\caption{
		A schematic diagram illustrating the full pipeline of the creation of mock galaxy catalogues (\textit{left column}) and the internal workflow of the Santa Cruz SAM (\textit{right}).
		On the left, we show the steps for lightcone construction from dark matter-only cosmological simulations to dark matter halo catalogues, labelled in grey. The semi-analytic modelling components that are responsible for creating the galaxy catalogues are labelled in red.
		On the right, we show a subset of internal processes in the Santa Cruz SAM that are most relevant to the predictions we present in this work.
		These processes are broken down into three groups: star formation related, black hole related, and observables. The physical processes are labelled with yellow boxes, blue boxes are for physical properties, and purple boxes are observable properties. The quantities with blue and purple labels are available in the mock galaxy catalogues.
	}
	\label{fig:SAM_FlowChart}
\end{figure*}

\begin{figure*}
	\includegraphics[width=2.4\columnwidth]{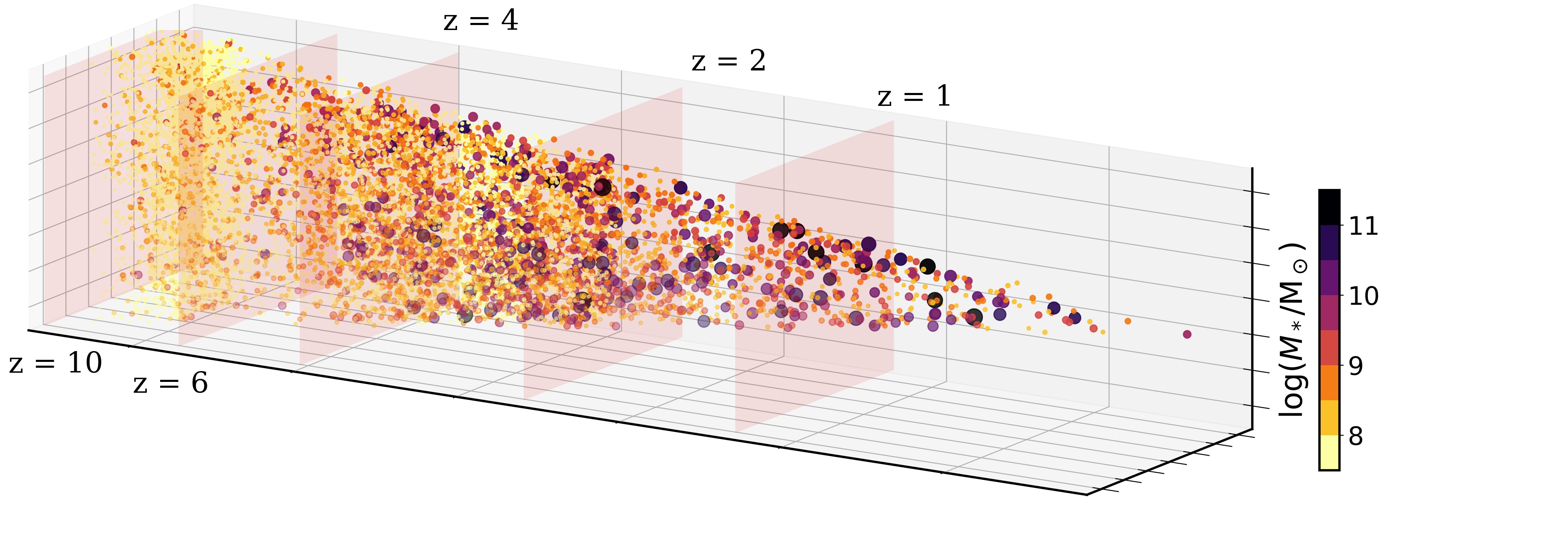}
	\caption{
		This is a three-dimensional visualization of a simulated lightcone of the COSMOS field. The cone-shaped feature is a manifestation of the predicted physical positions and distances for objects expected within the survey area along our line-of-sight (viewing from the right edge of the figure). Simulated lightcones of this sort offer a new perspective to understand and visualize the galaxies and quasars detected in deep-field surveys. 
		This plot only includes central galaxies.
		Galaxies in this plot are both colour-coded and size-coded by their stellar mass. To aid visibility in this plot, this plot only contains 1\% of objects with $\log(M_*/\text{M}_\odot)\geq8.0$ at $z \leq 2$, 2\% of objects with $\log(M_*/\text{M}_\odot)\geq7.5$ between $2 < z < 7$, and 25\% of objects with $\log(M_*/\text{M}_\odot)\geq7.0$ at $z \geq 7$ are uniformly sampled from the simulated lightcone.
	}
	\label{fig:lightcone_illustration}
\end{figure*}

\section{Lightcone Construction Pipeline with Physical Models}
\label{sec:models}

In this section, we provide a concise summary of the semi-analytic model (SAM) for galaxy formation developed by the Santa Cruz group \citep{Somerville1999, Somerville2008, Somerville2012, Somerville2015, Popping2014}. We refer the reader to these papers for a full description of the model components. The specific modelling and parameter choices for galaxy and AGN formation are documented in \citetalias{Yung2019} and \citetalias{Yung2021}. Free parameters in these models are calibrated as described in \citetalias{Yung2019} and \citet{Somerville2021}. Throughout this work, as well as the rest of the paper series, we adopt cosmological parameters $\Omega_\text{m} = 0.308$, $\Omega_\Lambda = 0.692$, $H_0 = 67.8$ \kms Mpc$^{-1}$, $\sigma_8$ = 0.831, and $n_s = 0.9665$; which are broadly consistent with the ones reported by the Planck Collaboration in 2015 (Planck Collaboration XIII \citeyear{Planck2016}). 
Throughout this work, all magnitudes are expressed in the AB system \citep{Oke1983} and all uses of log are base 10 unless otherwise specified. The calculations presented in this work make use of \texttt{ASTROPY} \citep{Robitaille2013,Price-Whelan2018}, \texttt{NUMPY} \citep{vanderWalt2011}, \texttt{SCIPY} \citep{Virtanen2020}, and \texttt{pandas} \citep{Reback2022}.

\subsection{Physical components of the galaxy formation model}

The SAM includes a fairly comprehensive set of physical processes, including gas cooling and accretion, star formation, stellar feedback, chemical evolution, black hole growth, and AGN feedback.
In this work, we adopt the fiducial `\citetalias{Gnedin2011}--\citetalias{Bigiel2008}2' model configuration that is consistent with the rest of the \emph{Semi-analytic forecasts for JWST} series papers. This configuration includes a multi-phase gas partitioning recipe motivated by numerical simulation results from \citet[][hereafter \citetalias{Gnedin2011}]{Gnedin2011}, which partitions the galactic disc into a neutral, ionized, and molecular component, and an observationally-motivated \molh-based star formation recipe from \citet[][hereafter \citetalias{Bigiel2008}]{Bigiel2008}, where the surface density of SFR, $\Sigma_\text{SFR}$, scales with the surface density of molecular hydrogen, $\Sigma_\text{\molh}$. Here we adopt a double power law SF relation, where the slope of the relation steepens above a critical density \citep[see][for details]{Somerville2015}. Stellar feedback is modelled by ejecting cold gas from the Interstellar Medium (ISM) at a rate proportional to the SFR, according to a phenomenological power-law function of the central halo circular velocity. Explorations of the dependence of the model predictions on the parameters of the star formation and stellar feedback recipes are presented in \citet{Yung2019} and \citet{Yung2019a}.
In Fig.~\ref{fig:SAM_FlowChart}, we show a schematic diagram that summarizes the internal workflow of the Santa Cruz SAM.

The free parameters in the galaxy model are calibrated to reproduce a variety of $z\sim 0$ observations, including stellar-to-halo mass ratio, stellar mass function, $M_{\rm BH}$--$M_\text{bulge}$ relation, stellar metallicity, and cold gas fraction (\citetalias{Yung2019} and \citetalias{Somerville2021}). 
The model outputs of the Santa Cruz SAM at low and intermediate redshift (e.g. $z \lesssim 6$) have been tested against observations in many works, including \citetalias{Somerville2015} and \citetalias{Somerville2021}, and at high redshift (e.g. $z \gtrsim 4$) in \citetalias{Yung2019} and \citetalias{Yung2019a}.
Note that in this paper series, as in all previous work with the Santa Cruz SAMs, the models have not been tuned to match observations at high redshift.

When a halo merges with another halo, the galaxy within the smaller halo becomes a `satellite' galaxy. The Santa Cruz SAM includes a semi-analytic treatment of the decay of satellite orbits due to dynamical friction, and of tidal stripping and destruction of satellites \citep[see][for details]{Somerville2008}. The radius of the satellite galaxy relative to the centre of the halo is tracked as the orbit decays; when a satellite reaches the centre it is merged with the central galaxy. In some cases, satellite galaxies can become completely tidally disrupted and destroyed. In this case their stars are added to a diffuse stellar halo, and their cold gas is added to the hot gas reservoir of the central halo.
This model was calibrated to match the sub-halo conditional mass function from the Bolshoi Planck N-body simulation \citep{Klypin2016}, but it does not predict the correct radial distribution of satellites within their host halos compared with Bolshoi Planck. Moreover, we found in this work that, when using the `native' satellite positions that are included in the published version of the SC SAM, we do not reproduce the observed clustering on small scales.
We therefore reassign the satellite positions in post-processing as follows. 

The positions of satellite galaxies are assigned in post-processing with an approach similar to the one detailed in \citet{Kakos2022}, which assumes that the satellites follow the same radial profile as the underlying dark matter halo, as described by an NFW model \citep*{Navarro1997}. 
The halo profile can be expressed in terms of the halo's virial mass as $M_\text{h} = M_\text{vir}\times u_\text{vir}(r)$, where
\begin{equation}
	u_{\rm vir} = \frac{f\left(   c_{\rm NFW}\,(r/R_{\rm vir})   \right)}{f(c_{\rm NFW})} 
\end{equation}
and $f \equiv \ln(1+x) - {x}/{1+x}$. 
Here $c_\text{NFW}$ and $R_\text{vir}$ are the NFW halo concentration parameter and virial radius of the (host) halo, respectively, which are obtained from the fits provided by \textsc{rockstar} (see \ref{sec:dm_catalogue} for details).
For each satellite galaxy, a random number $U_r$, which takes values between 0 and 1, is assigned, and the distance $r$ from the centre of the host halo is determined by solving $U_r - u_\text{vir}(r) = 0$.
Similarly, every satellite galaxy is assigned a random polar angle, $\theta \in [0,\pi)$,  and azimuthal angle, $\phi \in [0, 2\pi)$. 
The position of a satellite relative to the halo centre in Cartesian coordinates, $X'_{i}$, is given by
\begin{align}
	X'_1 &= r\sin(\theta)\,\cos(\phi)   \\
	X'_2 &= r\sin(\theta)\,\sin(\phi)   \\
	X'_3 &= r\cos(\theta) \text{.}
\end{align} 
These relative coordinates are added to their host halo position relative to lightcone axes $X_{i}$ to obtain their positions in the lightcone $X_{i,\text{lc}} = X_{i}+X'_{i}$. The physical positions on the sky $X_\text{sky} = R(X_\text{lc} - 0)$, where $R$ is the rotation matrix specific to the lightcone, are then converted to sky coordinates with
\begin{align}
	&{\rm RA }= \tan^{-1}\left(\frac{-X_{2,\rm sky}}{X_{1,\rm sky}}\right)\, \frac{180\deg}{\pi}\\
	&{\rm Dec}= \sin^{-1}\left(\frac{X_{3,\rm sky}}{ ||X|| }\right)\, \frac{180\deg}{\pi} \text{,}
\end{align}
where $||X|| = (X_\text{1,sky}^2 + X_\text{2,sky}^2 + X_\text{3,sky}^2)^{-2}$.

\subsection{Forward modelling to \emph{JWST} observables}
\label{sec:models_observables}
Based on the predicted star formation and chemical enrichment histories (SFHs, stored mass in bins of stellar age and metallicity), galaxies are assigned spectral energy distributions (SEDs) generated based on the results from the stellar population synthesis (SPS) model by  \citet{Bruzual2003}. The rest-frame SEDs are used to calculate rest-frame luminosities in filter bands as presented in the mock catalogue. In addition, quantities labelled with dust are calculated accounting for the effect of dust in the ISM. We assume a dust attenuation curve for starburst galaxies by \citet{Calzetti2000}. The $V$-band dust attenuation is calculated based on the surface density of cold gas and metallicity, based on a `slab' model as described in \citet{Somerville2012}, but adopting the latest recalibration of ISM dust optical depth presented in \citetalias{Yung2021}.
This update improves the agreement between UV LF predictions and observations at $z > 6$ compared to CANDELS DR1.

The rest-frame SEDs are then redshifted according to their redshift in the lightcone, and observed-frame magnitudes are computed, accounting for attenuation effects in the intervening IGM \citep{Madau1996}. Mid- to far-infrared bands also include additional contributions from dust emission, modelled as described in \citet{Somerville2012}, except that here we use the dust emission templates of \citet{Chary2001}. 
A new aspect of the lightcone catalogues described in this work is the addition of \emph{JWST}
NIRCam\footnote{\url{https://jwst-docs.stsci.edu/near-infrared-camera/nircam-instrumentation/nircam-filters}} broad- and medium-band photometry.

This dataset also include predicted photometry for the \emph{Roman Space Telescope} Wide Field Instrument (WFI), \emph{Euclid} visible imager (VIS) Near Infrared Spectrometer and Photometer (NISP-P), and \emph{Rubin Observatory}. We refer the reader to the companion \emph{Semi-analytic forecasts for Roman} paper for full descriptions \citep{Yung2022a}.

\begin{figure*}
	\includegraphics[width=2\columnwidth]{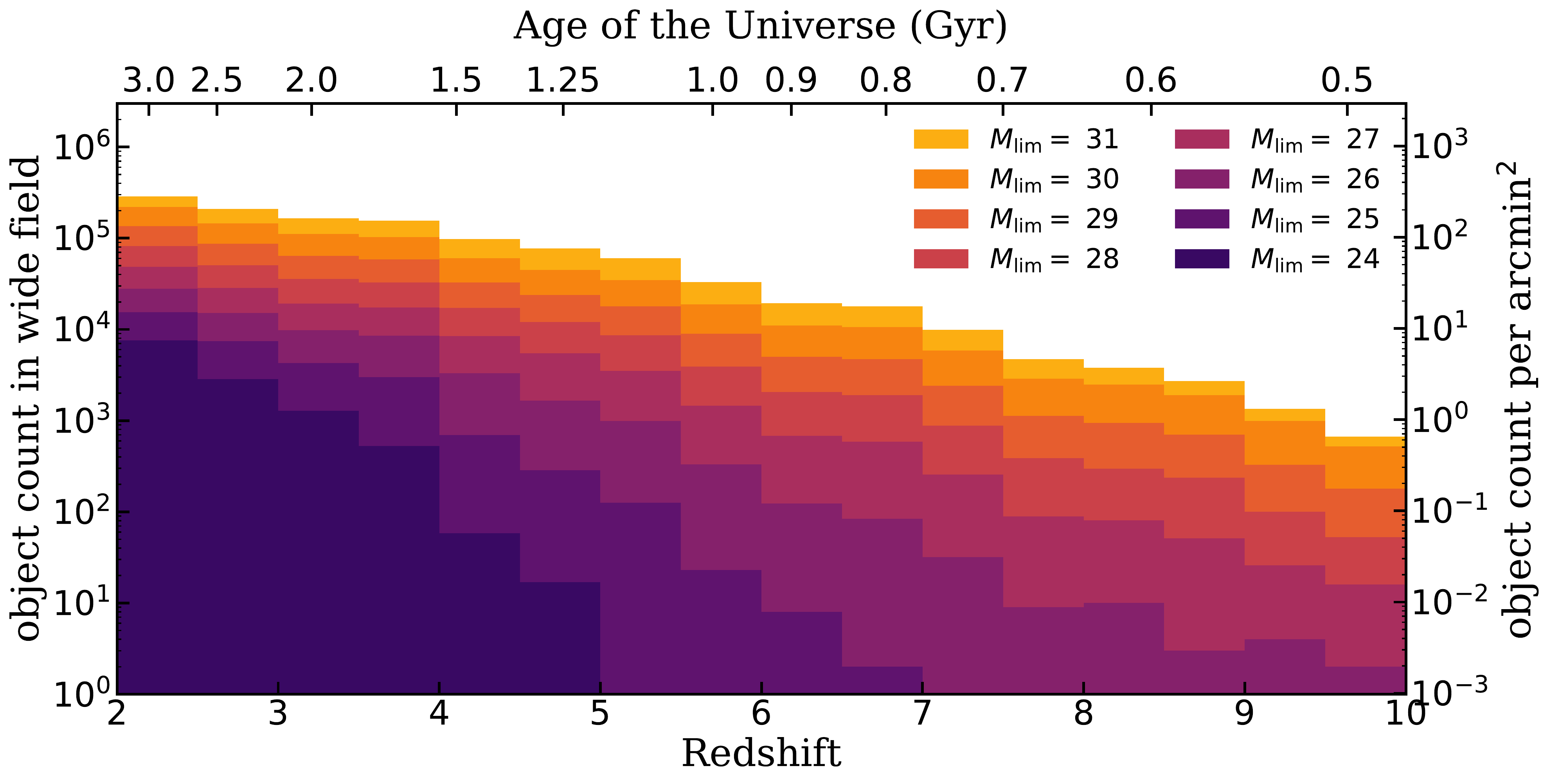}
	\caption{
		This histogram shows the number of objects predicted in a simulated wide-field lightcone (GOODS-N, realization 0) binned by redshift. This field spans 1024 arcmin$^2$. The number counts are colour-coded for a given limiting observed magnitude. On the right axis, we mark the equivalent number density of objects expected per arcmin$^2$, and on the top axis, we plot the age of the Universe equivalent to the marked redshift.
	}
	\label{fig:ObjCount_z}
\end{figure*}

\subsection{Dark matter halos catalogues and merger histories}
\label{sec:dm_catalogue}
Dark matter halos extracted from dark matter only cosmological N-body simulations serve as the the foundation of our simulated lightcones.
In this work, we present two sets of lightcones. The wide-field lightcones, which each span $\sim1000$\,\sqarcmin, is based on halos extracted from the Bolshoi-Planck simulation from the MultiDark suite \citep{Klypin2016}, which has a simulated volume of (250 Mpc $h^{-1}$)$^3$ and has a dark matter particle mass of $M_\text{DM}\sim1.5\times10^8$\Msun\;$h^{-1}$. 
This set of lightcones are the same as the ones from \citet{Somerville2021},
which have coordinates encompassing the five CANDELS legacy fields, with eight realizations of each field. We note that these wide-field lightcones are much larger than the actual observed CANDELS Fields. Future \emph{JWST} wide-field programs, such as COSMOS-Web in cycle 1, will be able to match, or even exceed, the area of these simulated wide-field lightcones.
In addition, we provide a set of eight realizations of deep-fled lightcones, each spanning 132\,\sqarcmin, constructed with halos extracted from the IllustrisTNG-100 dark matter-only simulation \citep{Nelson2019}, which has a simulated volume of (75 Mpc $h^{-1}$)$^3$ and a dark matter particle mass of $M_\text{DM}\sim8.9\times10^6$\Msun. 
Halos in both cosmological simulations are identified using \textsc{rockstar} and \textsc{consistent trees} \citep{Behroozi2013b, Behroozi2013c}. For more details on the halo catalogues, readers should refer to \citet{Rodriguez-Puebla2016} and \citet{Gabrielpillai2021}. As in those work, we adopt the virial mass definition from \citet{Bryan1998}.
Mock lightcones are then created using the \texttt{lightcone} package from \textsc{UniverseMachine} \citep{Behroozi2020}.

For each lightcone realization, the \texttt{lightcone} tool picks a random origin and viewing angle within the base dark matter-only simulation (see Table \ref{table:lightcone_specs}), and includes all halos that fall within the specified survey area. The tool makes use of the periodic boundary conditions when halos lie beyond the boundary of the simulated volume. 
The distance along the lightcone axis determines the redshift of the simulation snapshot from which halo properties are taken. 
While the lightcones in this paper were allowed to pass through the same region of the simulation volume multiple times, since the halos are sampled at a random angle, it is unlikely that a slice of the lightcone will be repeated in the same redshift slice (which happens only if halos are sampled in a slice that is perpendicular to the boundary of the simulation).

Using the virial mass of each halo in the lightcone as the `root mass', a Monte Carlo realization of the merger history of each halo were created using an extended Press-Schechter (EPS) based method \citep{Somerville1999a,Somerville2008}. Merger trees are resolved down to progenitors with a mass of at least a 100th of the root mass for all halos.
Merger trees are only generated for halos with $M_\text{h} > 10^{10}$\Msun\ and $M_\text{h} > 5.75 \times 10^{8}$\Msun\ for the wide-field and ultra-deep lightcones, respectively, which is equivalent to requiring at least 64 dark-matter particles for a halo to be resolved.
We note that at $z > 10$ for the ultra-deep lightcones, we only include halos with $M_\text{h} > 3 \times 10^{9}$\Msun\ to avoid processing low-mass halos in regimes where the merger tree algorithm become highly uncertain, and galaxies are likely to be too faint to be observed with \emph{JWST}.
The steps involved in constructing halo lightcones and merger trees are summarized in Fig.~\ref{fig:SAM_FlowChart}.

In Fig.~\ref{fig:lightcone_illustration}, we show an \emph{illustration} of the comoving spatial distribution of a subset of galaxies in one of the simulated lightcones for the COSMOS field.
In order to avoid saturation, this plot only contains 1\% of objects with $\log(M_*/\text{M}_\odot)\geq8.0$ at $z \leq 2$, 2\% of objects with $\log(M_*/\text{M}_\odot)\geq7.5$ between $2 < z < 7$, and 25\% of objects with $\log(M_*/\text{M}_\odot)\geq7.0$ at $z \geq 7$ that are uniformly sampled from the simulated lightcone.
Note that in this illustration, the data points are colour-coded and size-coded according to the predicted stellar mass of the simulated galaxies. The size of the data points does not reflect the galaxies' angular size or physical size. Satellite galaxies are omitted from this illustration.

In previous papers in the \textit{Semi-analytic forecasts for JWST} series, instead of a halo catalogue from an N-body simulation, we used a grid of halo masses, where each mass bin contained the same number of halo realizations. These results are then weighted by the halo mass function to obtain cosmologically representative quantities. This `grid mode' approach allows more efficient sampling of halos across an extremely wide mass range.  For instance, in \citetalias{Yung2019} through \citetalias{Yung2020}, we adopted a grid of one hundred halo masses equally spaced spanning $V_\text{vir} = 20$--500 \kms\ at output redshifts between $z = 4$ to 10 at integer increments.
In \citetalias{Yung2020a}, we extended these outputs with identical configurations up to $z = 15$. In order to better capture AGN powered by supermassive black holes found in massive, rare halos, in \citetalias{Yung2021} we adopted a new mass grid of two hundred masses spanning $V_\text{vir} = 100$--1400 \kms\ at discrete output redshifts between $z = 2$ to 7 at half-integer increments.
For each halo mass in the grid, one hundred Monte Carlo realizations of merger histories were created using the same EPS-based tree algorithm used here. 
The volume-averaged number density of these grid halos are weighted using halo mass functions from the Bolshoi-Planck simulations \citep{Klypin2016, Rodriguez-Puebla2016, Rodriguez-Puebla2017}.

The full object catalogues from these previous papers are also released via the same data portal as the lightcone data released with this work.

\begin{figure*}
	\includegraphics[width=2\columnwidth]{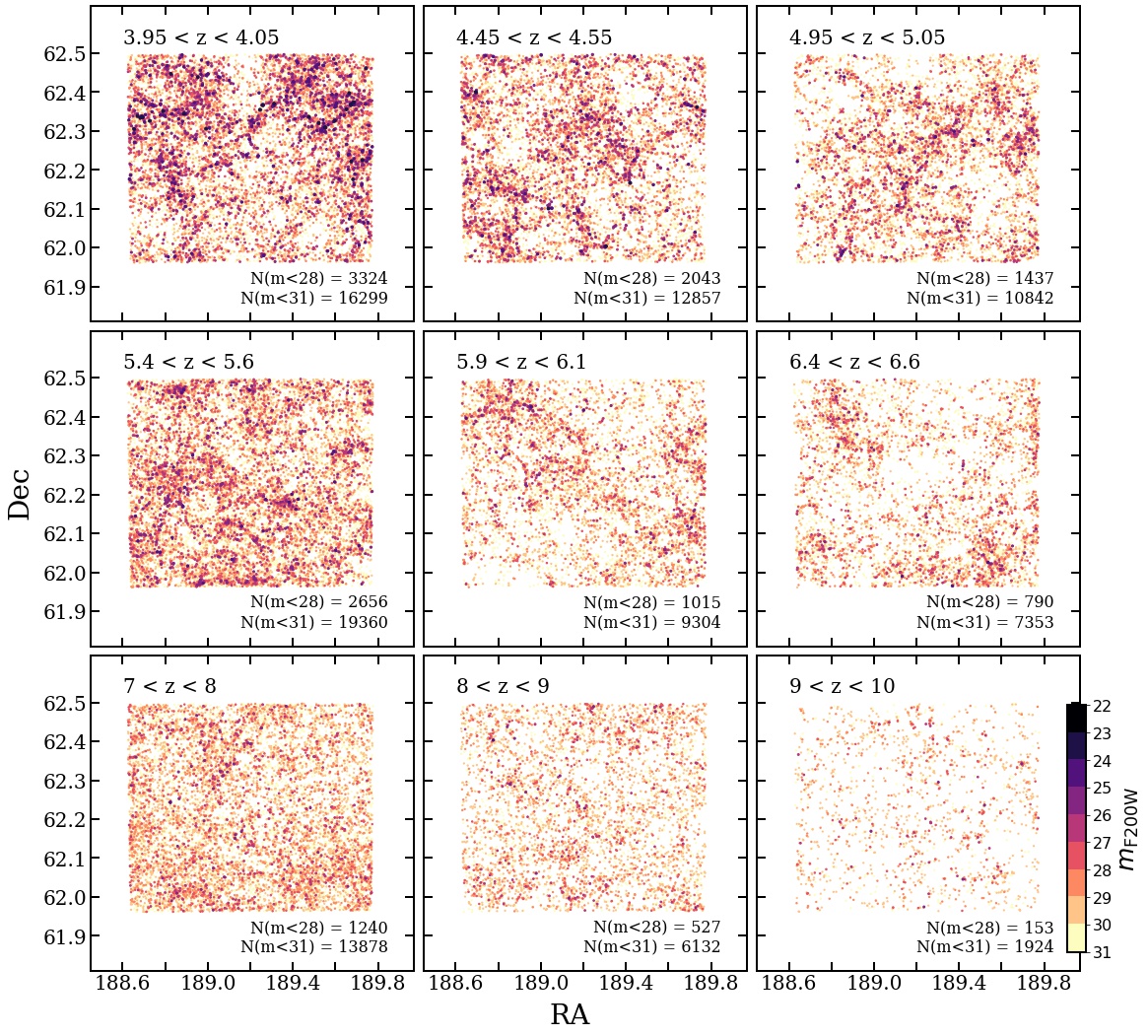}
	\caption{
		A summary of the footprint and galaxy populations of the first realization of the simulated GOODS-N lightcones ($1024$\,\sqarcmin) at various redshift slices between $z\sim 4$ to 10. The data points are colour-coded by the predicted observed-frame IR magnitudes in the NIRCam F200W band. The sizes of the data points are also scaled to emphasize brighter objects and do not reflect their predicted angular sizes. In addition, we show the counts of bright and faint objects within each panel with $m_\text{F200W} < 28$ and < 31, respectively.
	}
	\label{fig:footprint_F200W_EGS}
\end{figure*}

\begin{figure*}
	\includegraphics[width=2\columnwidth]{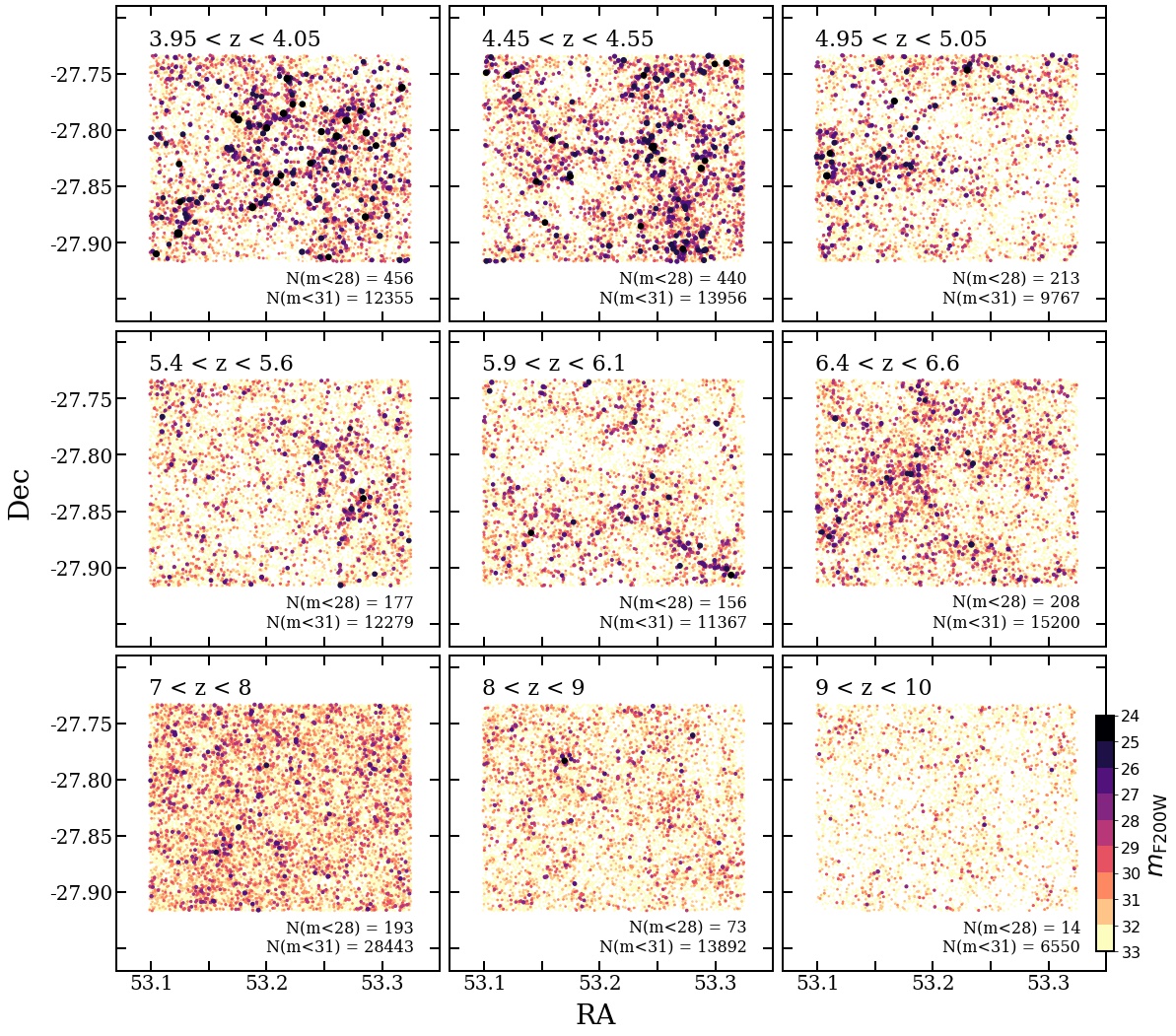}
	\caption{
		A summary of the footprint and galaxy populations of the first realization of the ultra-deep lightcones ($132$\,\sqarcmin) at various redshift slices between $z\sim 4$ to 10. The data points are colour-coded by the predicted observed-frame IR magnitudes in the NIRCam F200W band. The sizes of the data points are also scaled to emphasize brighter objects and do not reflect their predicted angular sizes. In addition, we show the counts of bright and faint objects within each panel with $m_\text{F200W} < 28$ and < 31, respectively.
	}
	\label{fig:footprint_F200W_DEEP}
\end{figure*}

\section{Simulated data products}
\label{sec:data_release}

Based on the physical models that have been examined extensively and shown to reproduce existing observational constraints up to $z \sim 10$, we present predictions for large populations of galaxies across wide ranges of redshift and halo masses, providing a comprehensive compilation of photometric and physical properties. In addition, we also provide full high-resolution spectra and star formation histories. 

\subsection{\emph{JWST} wide-field lightcones}
The five CANDELS fields are some of the most extensively surveyed patches of the sky and are expected to be frequently revisited by future surveys with \emph{JWST} and other instruments. 
Expanding on the same framework that generated the SC-SAM simulated lightcones presented in \citetalias{Somerville2021}, we provide additional \textit{JWST}-specific predictions, including infrared photometry for NIRCam broad- and medium-band filters, and include fainter galaxies that are expected to be detected by \emph{JWST} by relaxing the luminosity threshold from \emph{HST}/WFC3 $m_\text{F160W} = 29$ to 42. 
These lightcones include galaxies spanning $z=0$ to 10.
The simulated COSMOS, EGS, GOODS-N, GOODS-S, and UDS fields span 697, 782, 1024, 1599, and 1260\,\sqarcmin, respectively.
The modelling pipeline includes a hard stellar mass limit of $M_\text{*,lim} = 10^7$\Msun, such that galaxies with masses below this limit are not recorded. 
This threshold is set relative to $M_\text{h,res}$.
These wide-field lightcones contain galaxy samples that are complete across the range of rest-frame UV luminosity $-16 \lesssim M_\text{UV} \lesssim -22$.
We summarize some of the key specifications of these lightcones in Table~\ref{table:lightcone_specs}.
For the rest of this work, we present results from the simulated GOODS-N field to represent the wide-field lightcone results unless specified otherwise.
This field is chosen because its area is close to the average among the five wide fields and the aspect ratio is close to a square.

In Fig.~\ref{fig:ObjCount_z}, we present the object counts as a function of redshift using the first realization of the simulated GOODS-N lightcone.
The number of galaxies in this lightcone is broken down by the observed-IR magnitude in the NIRCam F200W broad-band filter $m_\text{F200W}$.
We also show the object counts normalized to the survey area on the vertical axis, and the corresponding age of the universe on the horizontal axis. 
On one hand, this figure illustrates the number of objects predicted in the simulated field as a function of redshift (left axis).
On the other hand, this figure can be used as an easy look-up table for the number of objects expected as a function of redshift, survey depth (or equivalently exposure time), and survey area (using the normalized object counts).

Fig.~\ref{fig:footprint_F200W_EGS} shows several cross-sections of the same lightcone.
These slices are taken perpendicular to the line of sight, spanning $z \sim 4$ to 10.
The object counts above two expected detection limits $m_{lim}=$ 28 and 31 are also shown.
The predicted galaxies in these figures are both size-coded and colour-coded by their observed-frame IR magnitude in the NIRCam broad-band F200W filter, $m_\text{F200W}$. This figure demonstrates the evolution of the number of objects and their spatial distribution as a function of redshift.

Figs.~\ref{fig:ObjCount_z} and \ref{fig:footprint_F200W_EGS} together provide a comprehensive view of the simulated lightcone, illustrating how the predicted galaxies are distributed on the sky and in the redshift direction. 
The lightcone construct of these mock galaxy catalogues enables a wide variety of predictions for observations, such as predictions that are sensitive to the survey geometry and quantities that are measured with non-negligible $\Delta z$, which are not achievable with a grid of halos (e.g. previous papers in this series) or with a simulated snapshot at a single redshift or a coarse grid of redshifts (e.g. most state-of-the-art cosmological simulations). 
Furthermore, the geometries and depths of these lightcones can be further customized to match real surveys, which enables direct comparison with an observed field, making them an ideal tool for survey planning and interpretation.

These wide-field lightcones reach a comparable depth to that expected for planned wide-field \emph{JWST} surveys, such as CEERS, PRIMER, and COSMOS-Web.
For example, CEERS has a five-sigma magnitude limit of $m_{5\sigma}\sim29.0$
We note that the CEERS survey is expected to observe the EGS field with \emph{JWST} \citep{Finkelstein2017}. The simulated EGS field has been used to aid the planning of CEERS and has been released as part of the project's pre-launch data products\footnote{CEERS Simulated Data Release v1: \url{https://ceers.github.io/sdr1.html\#catalogs}}, these catalogues are further processed into simulated NIRCam, MIRI, and NIRISS images (Bagley et al. in preparation).

Some model adjustments made in this work, particularly the recalibration of dust optical depth parameters based on the results from \citetalias{Yung2021}, are applied to the galaxies and predicted quantities that were previously presented in \citetalias{Somerville2021}. Therefore, this work also doubles as the second data release (DR2) for the mock CANDELS lightcones.

\begin{table}
	\centering
	\caption{This table summarizes the dimension, area, and key configurations for the lightcones release with this work and the sets of available data.}
	\label{table:lightcone_specs}
	\begin{tabular}{lccc}
		\hline
		Specification                      & Wide (GOODS-N)    & Ultra-deep      \\
		\hline
		Dimension (arcmin)                 & $32\times32$      & $12\times11$    \\
		Area (arcmin$^2$)                  & 1024              & 132             \\ 
		\hline
		Base simulation                    & Bolshoi-Planck    & TNG-100 DM      \\
		$\log M_\text{h,res}$/\Msun        & 10.00             & 8.76            \\
		\hline
		$\log M_\text{*,lim}/\text{\Msun}$ &7.00               & 5.76            \\
		$M_\text{UV}$ range                & $-16$ to $-22$    & $-12$ to $-21$  \\
		redshift range                     & $0<z\leq10$       & $0<z\lesssim12$ \\
		\hline
	\end{tabular}
\end{table}

\subsection{\emph{JWST} ultra-deep lightcones}

In addition to the set of wide-field lightcones enhanced with \emph{JWST} observables, we introduce a brand-new set of ultra-deep lightcones with superior mass resolution but for a smaller survey area.
These lightcones cover an area of 132\,\sqarcmin, containing galaxies spanning $0 < z \lesssim 12$.
At $z \leq 10$, our model is configured to output all galaxies with stellar mass above a limit of $M_\text{*,lim} = 5.7\times 10^5$\,\Msun, which is set relative to the mass at which the halos are resolved in the halo catalogue $M_\text{h,res} = 5.7\times 10^8$\,\Msun.
These ultra-deep lightcones contain galaxy samples that are complete across the range of rest-frame UV luminosity $-12 \lesssim M_\text{UV} \lesssim -21$, or down to $m_\text{F200W}\sim34$ at $z \lesssim 10$.
For dark matter halos at $z > 10$, only halos above $M_\text{h,res} = 3\times10^{9}$\,\Msun are processed, which give a complete galaxy sample down to
$m_\text{F200W}\sim32$, with the assumption that galaxies forming in halos below this mass are too faint to be detected by \emph{JWST}.
We also deliver eight realizations of this simulated field, consistent with the wide-field lightcones.
The key specifications of these lightcones are summarized in Table~\ref{table:lightcone_specs}.

Fig.~\ref{fig:footprint_F200W_DEEP} shows the survey footprint for the first realization of the ultra-deep lightcones. Data points are both colour-coded and size-coded by their observed IR magnitude.
These new lightcones are geared towards predictions for \textit{JWST} deep-field surveys, such as NGDEEP, which is expected to reach a limit of $m_{5\sigma}\sim30.7$. This lightcone was used in the planning of the NGDEEP survey \citep{Finkelstein2021a}.
Even though the performance of the SAM has been thoroughly examined and shown to reproduce a wide variety of existing observational constraints in \citetalias{Yung2019} and \citetalias{Yung2019a}, 
we note that the physical properties for the predicted galaxy populations at $z \gtrsim 10$ are poorly constrained due to the current lack of direct observations.
We refer the reader to \citetalias{Yung2020a}, where the rest-frame UV luminosity functions for $z = 11$ to 15 were first presented, for a thorough discussion of the limitations and uncertainties of the SAM in this regime.
These predictions will be tested more stringently as high-redshift observational constraints from \emph{JWST} become available.

\section{Results}
\label{sec:results}

In this section, we present key predictions at high redshift that are derived from the simulated lightcones. Specifically, we present several one-point distribution functions and two-point correlation functions. These specific results are selected to demonstrate the advantages of the mock lightcones, which include the spatial distribution and redshift distribution along the line of sight, with survey area and depth that mimics real high-redshift galaxy surveys. We also take advantage of the multiple realizations available to explore field-to-field variance.

\begin{figure}
	\includegraphics[width=\columnwidth]{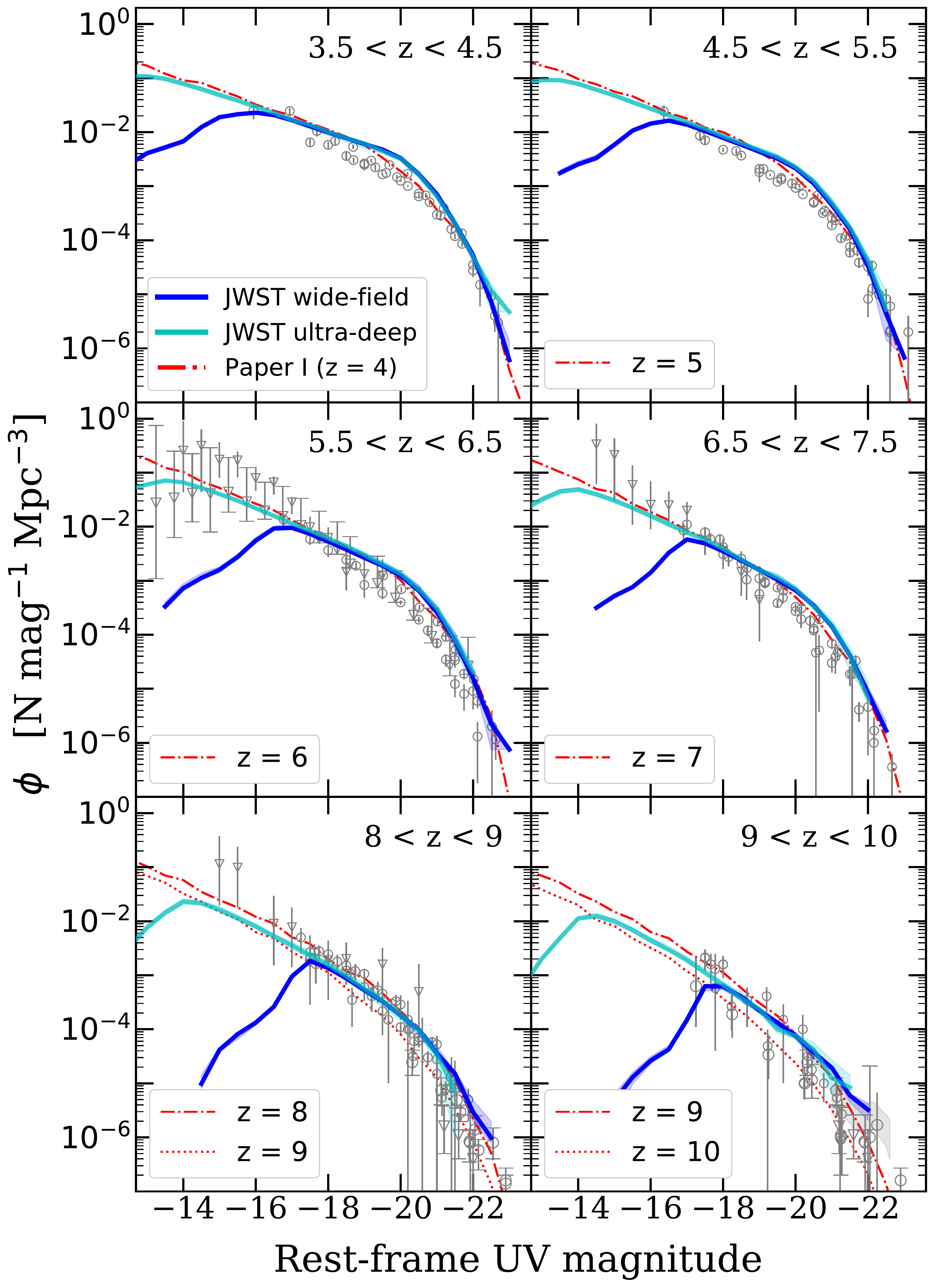}
	\caption{
		Predicted rest-frame UV LFs (including the effects of ISM dust attenuation and IGM extinction) at various redshift slices, together spanning the range between $z\sim4$ to 10. 
		The blue and cyan solid lines show results from the simulated \emph{JWST} wide-field (GOODS-N) and ultra-deep lightcones, respectively. The line shows the median of the eight realizations, and the shaded area marks the 16th and 84th percentile. We note that in most cases the shaded area is smaller than the line thickness.
		The red lines show results from a grid that is uniformly sampled in halo mass,  as presented in \citetalias{Yung2019}.
		We also show a compilation of observational constraints for comparison, where blank-field observations are shown by circle markers \citep{Finkelstein2015, Finkelstein2021, Bouwens2017, Bouwens2019, Atek2018, Oesch2018, Stefanon2019, Bowler2020} and results from lensed fields are shown with upside-down triangles \citep{Bouwens2017, Livermore2017}.
	}
	\label{fig:restUV_LFs}
\end{figure}

\begin{figure}
	\includegraphics[width=\columnwidth]{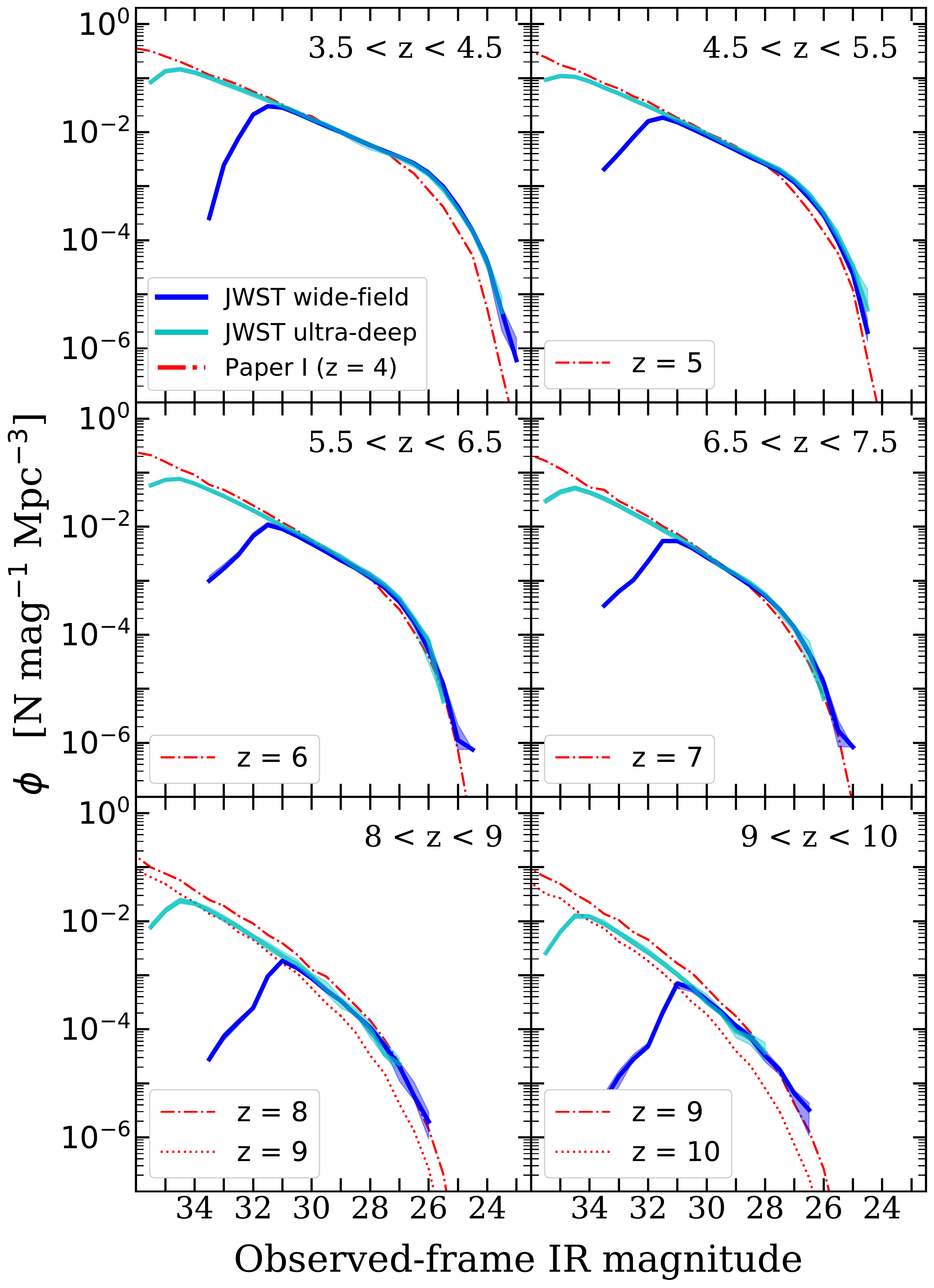}
	\caption{
		Predicted LFs for observed-frame magnitude in the NIRCam F200W filter (including the effects of ISM dust attenuation and IGM extinction) in various redshift slices, together spanning the range between $z\sim4$ to 10. 
		The blue and cyan solid lines show results from the simulated \emph{JWST} wide-field (GOODS-N) and ultra-deep lightcones, respectively. The line shows the median of the eight realizations, and the shaded area marks the 16th and 84th percentile. 
		We note that in most cases the shaded area is smaller than the line thickness.
		The red lines show results from a grid that is uniformly sampled in halo mass as presented in \citetalias{Yung2019}. The differences that can be seen between the lightcone and grid results are primarily due to evolution in galaxy properties and differences in the k-correction over the redshift range of the bin. The grid represents galaxy properties at a fixed `snapshot' in time. 
	}
	\label{fig:obsIR_LFs}
\end{figure}

\subsection{Luminosity functions}
One-point distribution functions are an effective way to summarize the statistical characteristics of specific observable or physical properties over a population of detected galaxies. 
In this subsection, we show the one-point distribution functions of luminosities and magnitudes for galaxies in various redshift slices in the two sets of simulated lightcones.

\subsubsection{Rest-frame UV luminosity}
The rest-frame UV magnitudes are computed for our model galaxies by integrating over synthetic stellar SEDs using a tophat filter centred at 1500\,\AA, with a full width at half maximum (FWHM) of 225\,\AA. 
These magnitudes include ISM dust attenuation as described in Section \ref{sec:models_observables}.
Far-UV magnitudes calculated with a similar filter centred at 2300\,\AA\ are also available as part of the full object catalogues.
In Fig.~\ref{fig:restUV_LFs}, we show the rest-frame UV luminosity functions (UV LFs) at 1500\,\AA\ for galaxies from eight selected redshift slices of the lightcones, spanning a wide redshift range from $z\sim4$ to $10$.
We show results from the wide-field (GOODS-N) and ultra-deep lightcones.
The UV LFs are first computed individually for each of the eight realizations.
We show the median in solid lines and the shaded regions mark the 16th and 84th percentile.

The turn-over on the faint end of the UV LFs is indicative of the decrease in completeness in our galaxy sample due to the mass resolution limit of the underlying dark matter halo catalogue (see Table \ref{table:lightcone_specs}).
On the other hand, the cut-off on the bright end of the luminosity functions reflects where the galaxy populations become too rare to produce statistically robust measurements for their volume-averaged number densities.
This also shows where \emph{JWST} wide surveys are not expected to constrain the bright end of the UV LFs as precisely as past large ground-based surveys. 
We note that while the wide-field and ultra-deep lightcones are made based on dark matter halo populations extracted from two different cosmological simulations, this does not have any noticeable effects on our predictions.

In addition, we show results based on a grid of haloes uniformly sampled in mass, as presented in \citetalias{Yung2019}, which covers halos spanning a mass range much wider than what can be resolved in a single typical numerical simulation (see Section \ref{sec:dm_catalogue}).
The \textit{volume-averaged} number densities of the halos in the grid are weighted by halo mass functions extracted from cosmological simulations. We note the excellent agreement between the two methods.

We compare these predictions to a compilation of observational constraints, including \citet{Finkelstein2015, Finkelstein2021, Livermore2017, Atek2018, Oesch2018, Bouwens2019, Stefanon2019, Bowler2020}.
Blank-field observations are shown by circle markers and the lensed results are shown with upside-down triangles.
We note that the cut-off in the bright end of the UV LFs is indicative of the limit where bright, rare objects are expected to be detected in these different types of surveys.
These high-redshift bright (massive) objects, which are often found by large surveys with ground-based instruments, are too rare for the expected survey volume of \emph{JWST}.
Here we highlight the objects that are expected to be detected in upcoming \textit{JWST} surveys, and contrast them with all objects predicted by our model.
We also note that lensed surveys will enable \textit{JWST} to detect objects with even fainter intrinsic magnitudes. See \citetalias{Yung2019} and \citet{Bouwens2022} for detailed comparison of the UV LF faint-end predictions and existing lensed survey constraints.

\subsubsection{Observed-frame near-IR magnitude}
In Fig.~\ref{fig:obsIR_LFs}, we show the distribution functions for observed-frame near-IR magnitude in the NIRCam broadband F200W filter, $m_\text{F200W}$, for all galaxies in the wide-field (GOODS-N) and ultra-deep lightcones. 
The calculation of observed-frame magnitudes include both ISM dust attenuation and IGM extinction (see Section \ref{sec:models_observables}). 
In addition, we show results based on a grid of halo mass as presented in \citetalias{Yung2019}, which covers haloes spanning a mass range much wider than a single typical numerical simulations (see Section \ref{sec:dm_catalogue}). 
Similar to the previous figure, this figure provides a quick assessment for the observed-frame IR luminosity range the galaxies span in these wide-field and deep-deep lightcones. We note that here, the agreement between the grid mode and lightcone predictions is not as good in some cases. This is due to the $k$-correction effects, which are more accurately modelled in the lightcone predictions presented here because the lightcones contain a continuous distribution of redshifts sampling the backwards past lightcone, while in our previous work, all galaxies were assumed to be at a set of discrete redshifts.

\begin{figure*}
	\includegraphics[width=1.5\columnwidth]{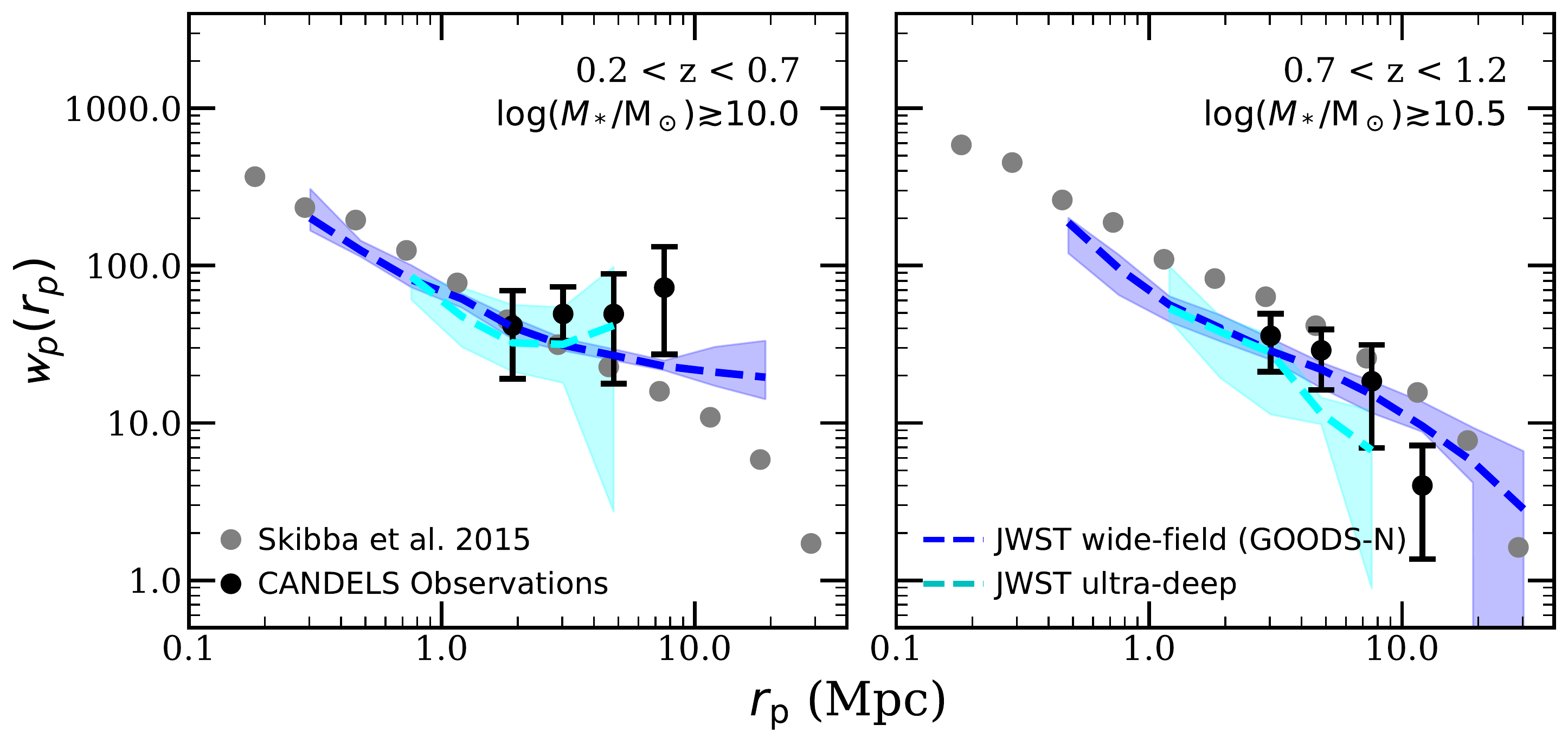}
	\caption{
		A comparison of the projected 2PCF, $\mathit{w}_{p}(r_p)$, between our calculations based on our mock lightcones and observations from PRIMUS and DEEP2 observations \citep{Skibba2015} in two redshift bins, $0.2<z<0.7$ and $0.7 < z < 1.2$, for galaxies with stellar mass $\log(M_* / \text{M}_\odot) \gtrsim 10.0$ and $\gtrsim 10.5$, respectively.
		We also show results for the 2PCF which we have computed from the five CANDELS observed fields, for the same stellar mass and redshift bins.
		For comparison, we show results from our wide-field (GOODS-N) and ultra-deep lightcones in blue and cyan, respectively.
		The dashed lines show the median of the eight realizations, and the boundaries of the shaded regions mark the 16th and 84th percentile over the different realizations.
	}
	\label{fig:low-z_clustering}
\end{figure*}

\begin{figure*}
	\includegraphics[width=1.5\columnwidth]{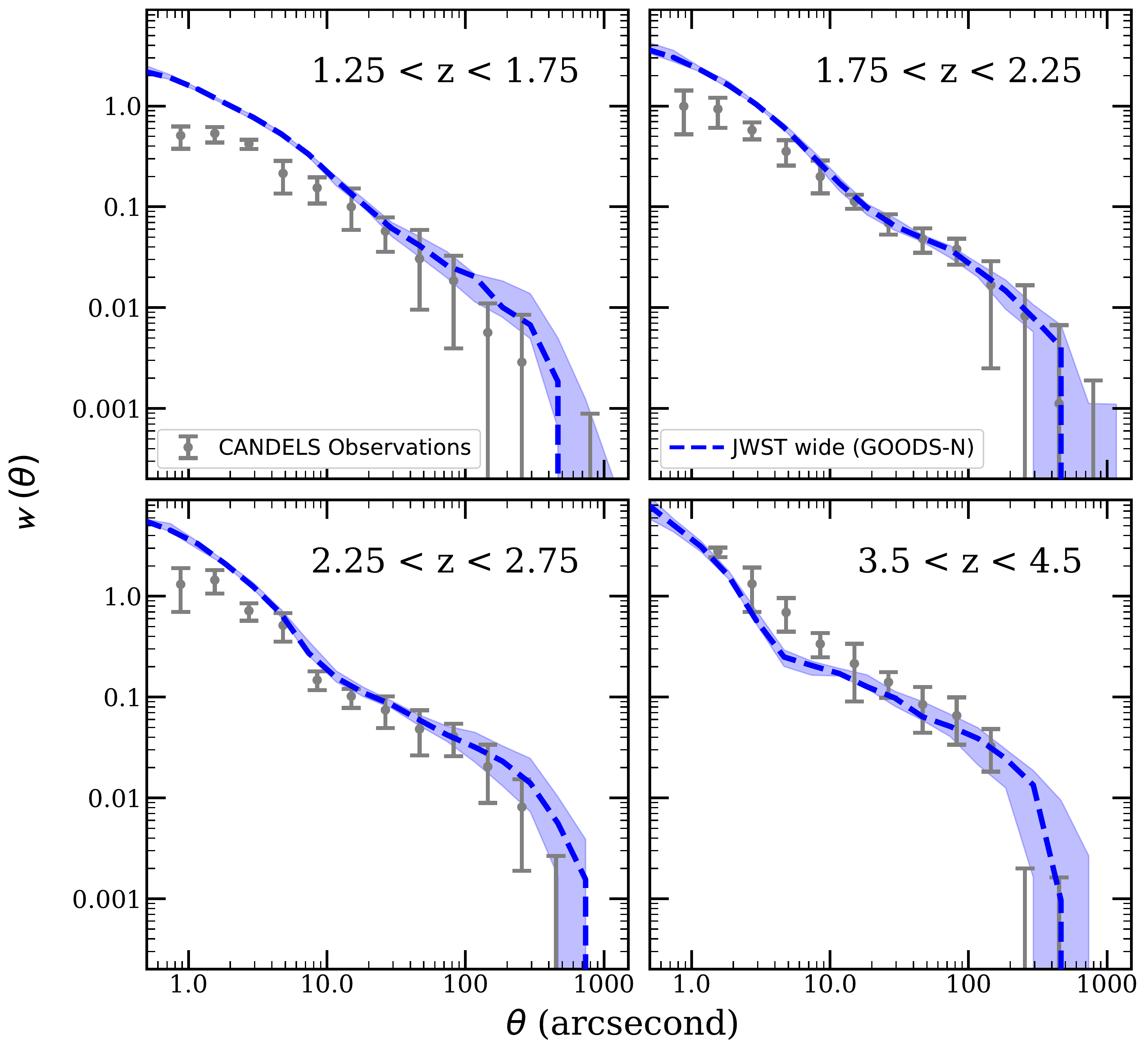}
	\caption{
		A comparison of the angular 2PCF, $\mathit{w}$($\theta$), at low and intermediate redshifts between the simulated lightcones and the CANDELS observational catalogues. A magnitude cutoff of $m_\text{F160W} \sim 26.5$ is applied to both sets of data.
		The dashed and shaded area show the median and the 16th and 84th percentile among the eight realizations of the simulated wide-field (GOODS-N) lightcones. The data point and error bars mark the median and 16th and 84th percentiles among the five CANDELS legacy fields.		
	}
	\label{fig:mid-z_clustering}
\end{figure*}

\subsection{Galaxy clustering statistics}
The two-point correlation function (2PCF) is a ubiquitous tool to characterize the spatial clustering of galaxies across different length scales \citep{Peebles1980}. 
In particular, the angular correlation function, $\mathit{w}(\theta)$, represents the angular separation, $\theta$, between random pairs of objects on the sky. 
A related quantity is the projected correlation function, $\mathit{w}_p(r_p)$, which is the correlation function in terms of the projected distance on the plane of the sky, integrated over a specific range in line-of-sight velocities.

In this subsection, we present selected 2PCFs for galaxies in our simulated lightcones. Lightcones are well suited for this application because, especially for high redshift clustering studies, it is often necessary to include galaxies from a relatively large bin in redshift in the 2PCF analysis. Galaxy properties may evolve over the time represented by the highest and lowest redshift edge of the bin. This effect is not included in theoretical predictions of the 2PCF from a single redshift slice. We make our clustering predictions accounting for the field geometry and spatial distribution of galaxies in our simulated lightcones, as well as the redshift distribution along the line of sight in a way similar to the three-dimensional distribution of objects in an \textit{observed} field.
The spatial locations of galaxies are determined by the physical coordinates of their host halos, which are inherited from the underlying N-body simulations. The contribution from peculiar velocities along the line of sight is included in the redshifts. The positions of satellite galaxies within the host halos are assigned as described in Section \ref{sec:models}. 
We calculate the 2PCFs using \textsc{corrfunc}\footnote{\url{https://github.com/manodeep/Corrfunc/}, v2.3.4} \citep{Sinha2020}, which utilizes the Landy-Szalay estimator \citep{Landy1993}.

\subsubsection{Low redshift}

In Fig.~\ref{fig:low-z_clustering}, we compare the projected 2PCFs from the wide-field (GOODS-N) and ultra-deep lightcones to measurements from the PRIMUS \citep{Coil2011, Cool2013} and DEEP2 \citep{Newman2013} galaxy surveys presented by \citep{Skibba2015}.
The comparison is done in two redshift bins, $0.2 < z < 0.7$ and $0.7 < z < 1.2$, for galaxies with stellar mass $\log(M_* / \text{M}_\odot) \gtrsim 10.0$ and $\gtrsim 10.5$, respectively.
These projected 2PCFs are measured by integrating $\xi(r_p,\pi)$ out to line-of-sight seperations of $\pi_\text{max} = 80 h^{-1}$ Mpc and $20 h^{-1}$ Mpc for PRIMUS and DEEP2, respectively.

We show results from the eight GOODS-N wide-field lightcones and the eight ultra-deep lightcones.
We computed the 2PCFs independently for each realization of the fields. These predicted 2PCFs are integrated out to $\pi_\text{max} = 40 h^{-1}$ Mpc.
The dashed line shows the median of the eight realizations and the shaded regions show the 16th and 84th percentile, characterizing the field-to-field variance. Note that the shaded regions are \emph{not} expected to be representative of the field-to-field variance in the observational surveys, as the areas and geometries of the mock and real surveys are very different.

In addition, we present new measurements of the 2PCF from the five observed CANDELS legacy fields, utilizing the theory-friendly catalogues from the CANDELS surveys curated by \citetalias{Somerville2021}.
The theory friendly catalogues are a set of catalogues for the five CANDELS field that provide a standardized set of photometric and physical properties for galaxies, including redshifts and stellar masses.
We utilize the setup with \textsc{corrfunc} same as the one applied to the mock lightcones, and calculate $w_p(r_p)$ for observed galaxies.
In this calculation, a random catalogue with matching object counts in generated  within an area similar to the observed fields, which is approximated with a quadrilateral shape that traces the edges of the field.
The projected auto-correlation functions are first calculated independently for each of the five observed CANDELS fields. 
In Fig.~\ref{fig:low-z_clustering}, the data points and error bars show the median and the 16th and 84th percentiles across the five fields.
We applied matching stellar mass cuts to the observed galaxy samples based on the inferred median stellar mass at observation time included in the theory-friendly catalogue, which is based on \citet{Pacifici2012, Pacifici2016}.

We note that the 2PCF estimate and the field-to-field variance can be sensitive to survey geometry, depth, and completeness, and redshift accuracy. Given that there are many differences between the PRIMUS and DEEP2 surveys and CANDELS, and that we have made the crude approximation that the depth of the CANDELS fields is uniform over their entire area, we find the agreement encouraging. For example, PRIMUS and DEEP2 utilize prism-based and spectroscopic redshifts, which are much more accurate than the CANDELS photometric redshifts that are all that is available for most galaxies. In addition, PRIMUS and DEEP2 have much larger areas than CANDELS. The `up-tick' in $w_p(r_p)$ at large separations is likely due to the small area of the CANDELS fields.

Overall, the SC SAM model predictions are in good agreement with observational estimates of the 2PCF. This is a genuine prediction of the models, as no information about galaxy clustering was used to calibrate the models. These results also demonstrate the range of pair separations over which we expect to be able to obtain robust measurements. This depends on survey area and depth, and object selection criteria. The theoretical results shown here are not intended to be representative of a particular observational survey, but users may download our catalogues and create customized predictions for a desired set of survey characteristics. We note that we are unable to generate meaningful projected 2PCF for the ultra-deep cone in the lowest redshift bin due to both the small physical scale the cone spans and the small number of galaxies that fall within the selection criteria.

\subsubsection{Intermediate redshift}
In Fig.~\ref{fig:mid-z_clustering}, we compare the angular 2PCFs from the wide-field lightcones with observations at intermediate redshift $1.25 \lesssim z \lesssim 4.5$.
The observational constraints are computed from the galaxies in the CANDELS theory-friendly catalogues as described in the previous subsection.
We select galaxies in both the theory-friendly catalogues and our simulated lightcones to be brighter than a magnitude limit of $m_\text{F160W} = 26.5$. The CANDELS wide fields are highly complete at this limit, and the photometric redshift estimates are relatively robust \citep[see][and references therein]{Somerville2021}.

We find qualitatively good agreement between the predicted and observed angular correlation functions at intermediate redshifts, at least on large scales.
We note that the wide-field lightcones are significantly larger than the observed fields. We refer the reader to \citetalias{Somerville2021} for detailed specifications for both datasets. 
We note that the 2PCF on small scales is dominated by galaxies that reside within the same host halo (sometimes called the `one-halo term'; the positions of these satellite galaxies are particularly uncertain in the SAM, so these small scale measurements should be given a lower weight in assessing the success of the model.
We also note that we have not attempted to add photometric noise or selection effects beyond a magnitude limit, or to simulate redshift uncertainties in the mock lightcones in the comparison presented here. The considerable uncertainties in the photometric redshifts that are predominantly available in CANDELS could have a significant effect on clustering measurements. However, this is intended to be a first order comparison to give us confidence that our predicted clustering properties are reasonable. We defer a more detailed comparison with observations to a future work.

\begin{figure*}
	\includegraphics[width=2.0\columnwidth]{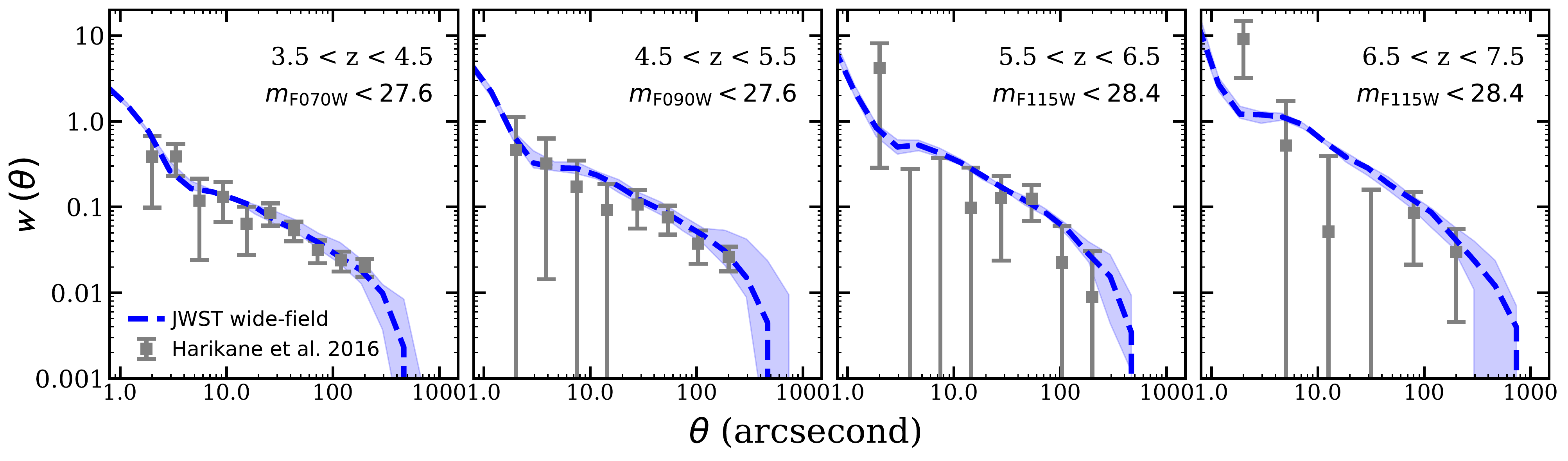}
	\caption{
		A comparison of the angular 2PCF, $\mathit{w}$($\theta$), at high redshifts between the simulated lightcones and observational constraints from \citet{Harikane2016}. A magnitude cut-off is applied to the simulated data to match the observations. In this comparison, we adopt the observed-IR magnitude in the NIRCam filter that is closest to rest-frame UV 1500\AA\ at the centre of the redshift bins, which is approximately equivalent to the selection criteria adopted for the observational constraints.
	}
	\label{fig:high-z_clustering_compare}
\end{figure*}

\begin{figure*}
	\includegraphics[width=1.85\columnwidth]{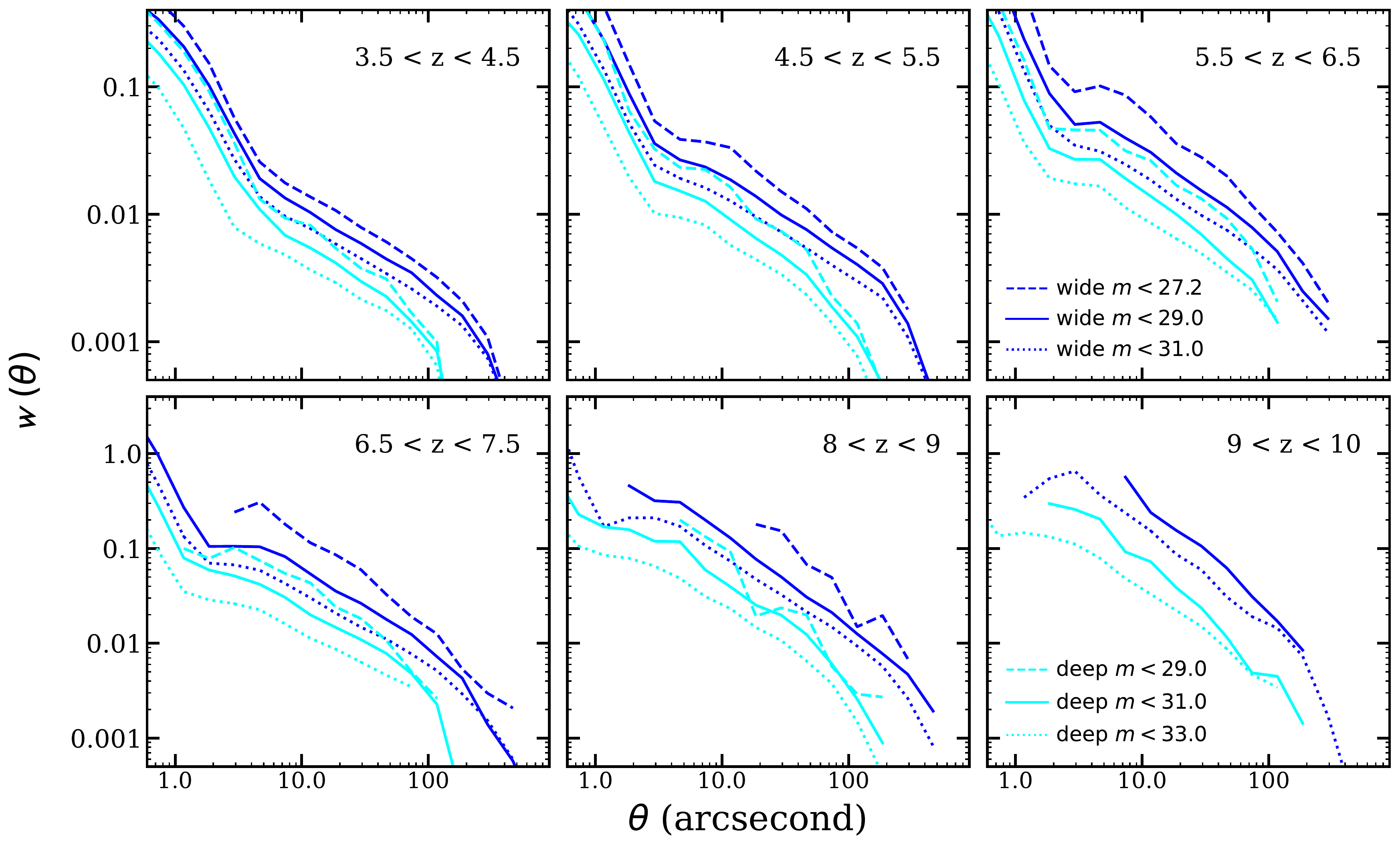}
	\caption{
		Predicted angular 2PCF, $\mathit{w}$($\theta$), at $z\sim4$ to 10. The blue and cyan lines show predictions from the wide-field and ultra-deep lightcones, respectively. These predictions adopted an observed-frame magnitude limit of $m_\text{F200W,lim} = 29$ and $31$, which are comparable to anticipated \emph{JWST} wide and deep surveys.
	}
	\label{fig:high-z_clustering_forecasts}
\end{figure*}

\begin{figure*}
	\includegraphics[width=1.85\columnwidth]{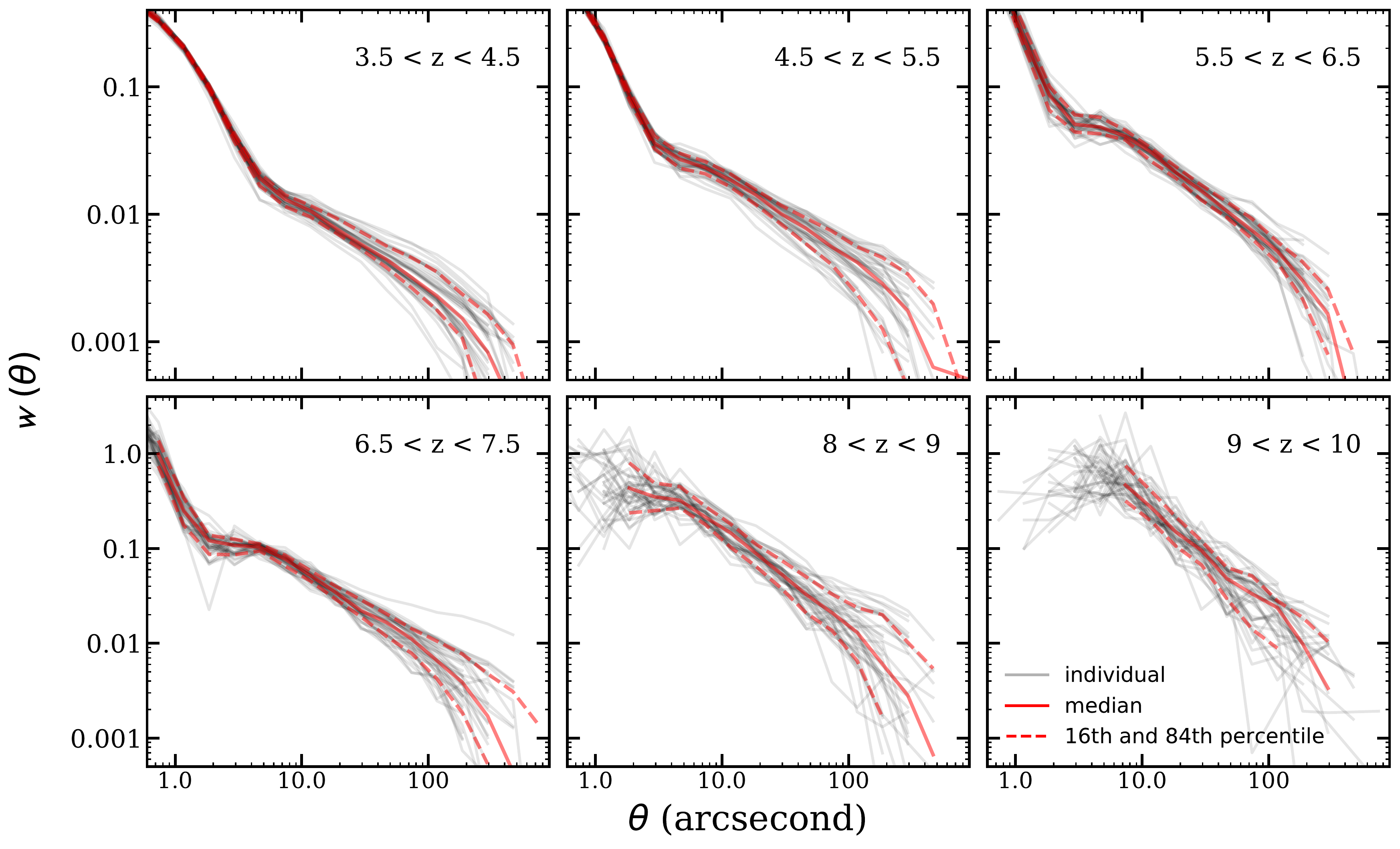}
	\caption{
		Predicted angular 2PCF, $\mathit{w}$($\theta$), at $z\sim4$ to 10 for galaxies $m_\text{F200W} < 29$ (same survey  configurations and limits as blue solid lines in Fig.~\ref{fig:high-z_clustering_forecasts}). The 2PCFs from the full set of 40 wide-field lightcones are shown individually in grey, and the median and 16th and 84th percentiles with solid and dashed red lines, respectively. Tabulated values of the predicted ACF are available in Table \ref{table:angular}.
	}
	\label{fig:high-z_clustering_scatter}
\end{figure*}

\subsubsection{High redshift}

In Fig.~\ref{fig:high-z_clustering_compare}, we compare the 2PCF from our simulated lightcones between $z \sim 4$--7 to observational measurements from \emph{Hubble} legacy deep imaging and Subaru/Hyper Suprime-Cam data presented by \citet{Harikane2016}. 
These observed results are based on over $\sim10,000$ Lyman-break galaxies identified in 10 deep optical-near-IR imaging data sets over the five CANDELS fields and the Hubble Frontier Field parallel fields. 
For 2PCFs calculated with our simulation lightcones, we apply a matching magnitude limit $m_\text{UV} < 27.6$ at $z\sim4$ and $\sim5$, and $m_\text{UV} < 28.4$ at $z\sim6$ and $\sim7$.
The observed-frame magnitude limit is applied to the filter band with central wavelength closest to rest-frame 1500\,\AA in the given redshift range. This is also individually marked in the panels of Fig.~\ref{fig:high-z_clustering_compare}.
The redshift bin of the observed sources is estimated based on the colour selection criteria detailed in \citet{Bouwens2015}.
In this comparison, we adopt coarse redshift bins for our sample to compensate for the large redshift uncertainties in the observed samples. Similar results are also reported by \citet{Barone-Nugent2014} and \citet{Qiu2018}. However, 2PCF calculations are highly sensitive to sample selection criteria (e.g. by stellar mass, by rest-frame luminosity, or by observed-frame luminosity), so it is difficult to consolidate the results in a homogeneous way. We find strikingly good agreement between our model predictions and the observed clustering properties of these high redshift galaxies reported by \citet{Harikane2016}, within the large observational error bars.

Fig.~\ref{fig:high-z_clustering_forecasts} shows \emph{forecasts} for 2PCF that may be measured by \emph{JWST}. The predictions for wide surveys utilise the full 1024\,\sqarcmin\ and are marked with blue lines, with the solid line marking the expected survey depth of a typical \emph{JWST} wide-field survey of $m_\text{F200W} < 29.0$. This prediction is accompanied by bracketing cases of a shallower limit $m_\text{F200W} < 27.2$ and an ultra-deep limit $m_\text{F200W} < 31.0$. Similarly, we show predictions for deep surveys over the 132\,\sqarcmin\ field. The assumed deep survey depth $m_\text{F200W,lim} \sim 31$ is represented by the solid lines, and is accompanied by results with shallower $m_\text{F200W} < 29.0$ and $m_\text{F200W} < 33.0$ depths. We note that at $9 < z < 10$, the number of bright objects, $m_\text{F200W} < 27.2$ and $< 29.0$ in the wide and deep fields, respectively, are not sufficient for a meaningful 2PCF calculation. All lines shown in this comparison are the medians over eight realizations calculated as described in previous sections. We emphasize again that the area and volume of the mocks is significantly larger than the anticipated area/volume of \emph{JWST} surveys, at least those that will be available in the first several years of operation. Over time, we may be able to expand the surveyed area to be comparable to that of our mock surveys.

\begin{figure*}
	\includegraphics[width=1.70\columnwidth]{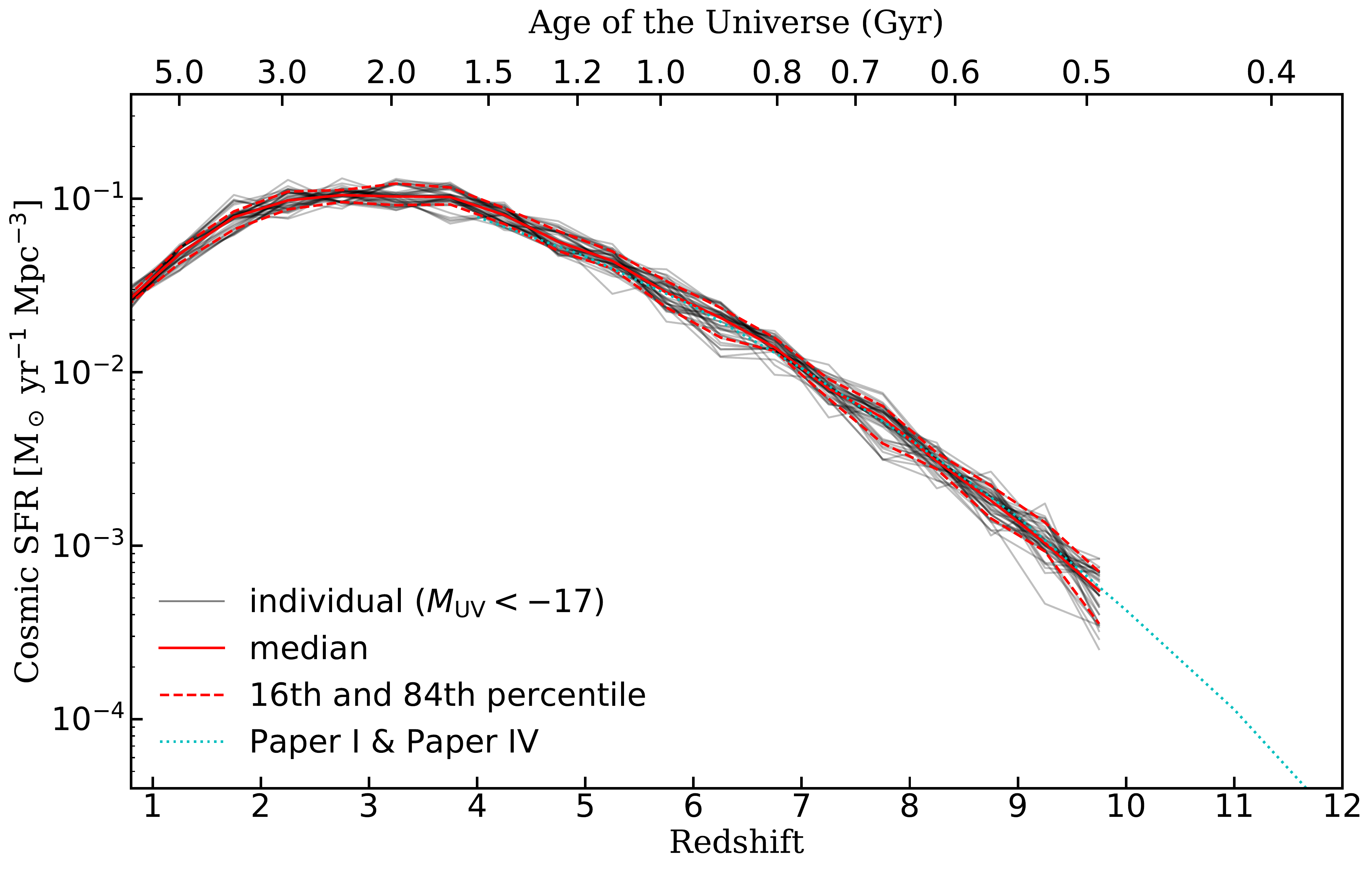}
	\caption{
		Predicted cosmic SFR density (SFRD) as a function of redshift at $z\sim 0$ to 10 from the wide-field lightcones for galaxies with rest-frame $M_\text{UV} < -17$. We also show the age of the Universe equivalent to the marked redshifts. The cosmic SFRD from the full set of 40 lightcones are shown individually in grey, and the median and 16th and 84th percentiles with solid and dashed red lines, respectively. In addition, we show the calculation from \citetalias{Yung2019} ($z = 4$ to 10) and \citetalias{Yung2020a} ($z = 11$ to 15). Tabulated values of the predicted cosmic SFR are provided in Table \ref{table:csfr}.
	}
	\label{fig:CSFR_all_fields}
\end{figure*}

\subsection{Field to field variance}

In this section, we leverage the combined set of 40 simulated wide-field lightcones (the ones overlapping with EGS, COSMOS, GOODS-N, GOODS-S, and UDS fields) to demonstrate the potential spread in the 2PCFs and cosmic SFR due to field-to-field variance expected in future \emph{JWST} wide-field surveys of similar sizes.
Fig.~\ref{fig:high-z_clustering_scatter} shows the field-to-field variation in the predicted wide-field 2PCFs with $m_\text{F200W} < 29.0$ (same as the solid blue lines in Fig.~\ref{fig:high-z_clustering_forecasts}). 
We show the 2PCF from individual fields in grey. We mark the median and 16th and 84th percentiles with solid and dashed red lines, respectively. 
Tabulated values of the predicted ACF are available in Table \ref{table:angular}. Additional ACFs computed with galaxies from the ultra-deep lightcones down to $m_\text{F200W} \sim 31$ are available in Table \ref{table:angular2}.

Combining results in Figs.~\ref{fig:high-z_clustering_forecasts} and \ref{fig:high-z_clustering_scatter}, we expect that \emph{JWST} will be able to improve constraints on galaxy clustering statistics for separation scales up to $\sim 10$ arcseconds up to $z\sim 7$. At higher redshifts, the number of bright galaxies decreases significantly and the statistics are significantly impacted by the number of sources in the limited survey volume.
We also show that the expected field-to-field variance in angular correlation function at separation scale $\sim100$ arcseconds can be up to a factor of 2 for galaxies with $m_\text{F200W} < 29$ in fields with area $\sim 1000$\,\sqarcmin.
The effect of survey area on galaxy clustering measurements will be explored more systematically and quantitatively in \citet{Yung2022a}.

Similar to the experiment shown in the previous figure, Fig.~\ref{fig:CSFR_all_fields} shows the cosmic SFR density (SFRD) as a function of redshift between $z \sim 0$ to 10 predicted in all 40 wide-field lightcones.
The cosmic SFRD is calculated by summing over the contributions from galaxies with $M_\text{UV} < -17$ and normalizing by the comoving volume in redshift bins with width of $\Delta z=0.5$.
This shows that the cosmic SFRD estimated from one wide field (of the order of $\sim 1000$\,\sqdeg, which is a few times larger than an observed CANDELS field) will be able to constrain the cosmic SFRD down to $\sim5\%$ with a field-to-field variance uncertainty of $\sim10\%$ to $35\%$ from $z = 4$ to 10 for galaxies with $M_\text{UV} < -17$.
In addition, we show the calculation from \citetalias{Yung2019} ($z = 4$ to 10) and \citetalias{Yung2020a} ($z = 11$ to 15). 
These predictions are made with the more flexible `grid mode' for halo across a wide mass range (see Section \ref{sec:dm_catalogue}).
Tabulated values of the predicted cosmic SFR are provided in Table \ref{table:csfr}.

Field-to-field variance can be one of the major sources of uncertainty in estimates of the number of objects from observational studies. 
\citet{Somerville2004} and \citet{Moster2011} have provided an analytic `cookbook' for cosmic variance for stellar mass selected galaxies in CANDELS-sized deep pencil beam surveys (of the order of hundreds of \sqarcmin) for $z \sim 0.5$ to 4, based on an empirically established halo occupation distribution (HOD) model.
In this work, we calculate the field-to-field variance for magnitude or luminosity selected galaxies at high redshift utilizing the full set of 40 realizations of wide-field lightcones, which span $\sim1000$\sqarcmin\ each.

\begin{figure*}
	\includegraphics[width=1.65\columnwidth]{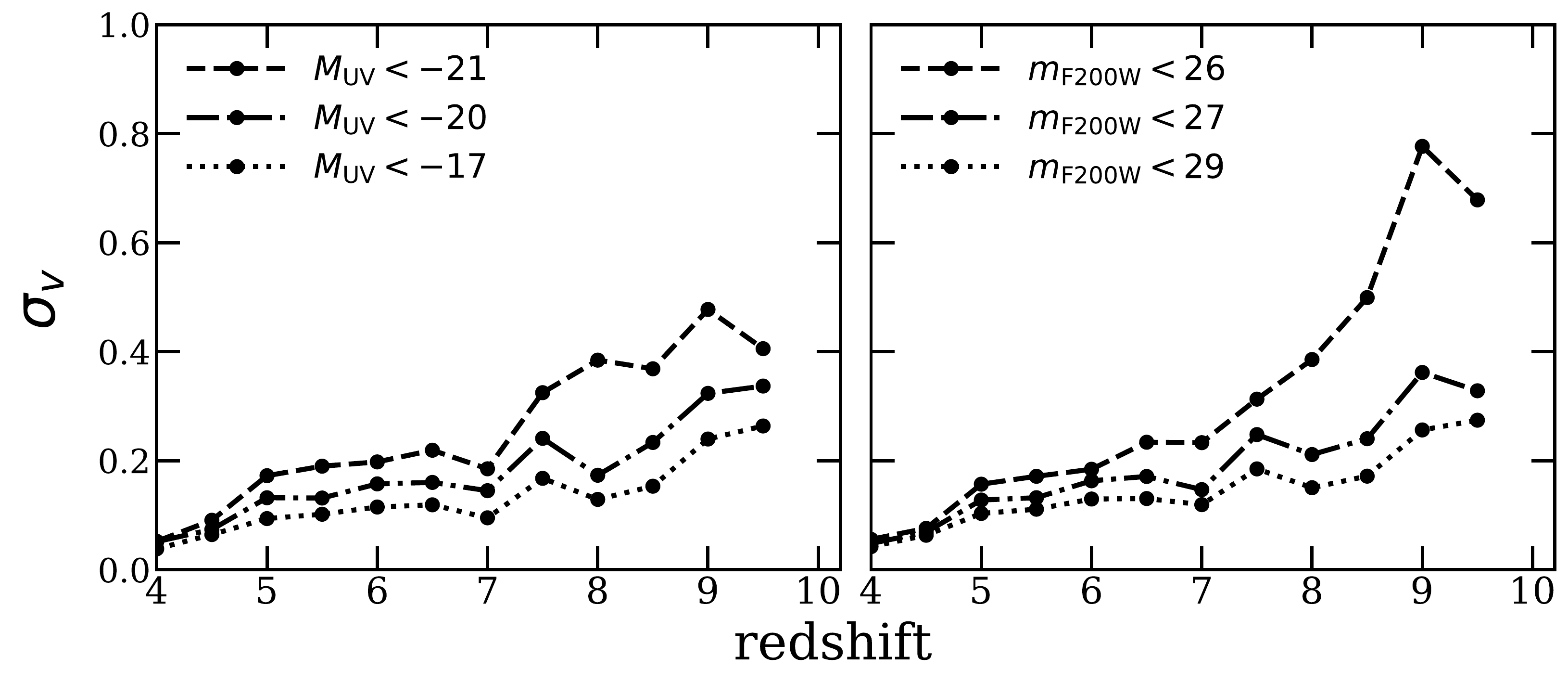}
	\caption{
		Square root of field to field variance, $\sigma_\text{V}$, as a function of redshift between $4 < z < 10$ calculated for galaxies across the full set of 40 realizations of wide-field lightcones, each spanning $\sim 1000$\sqarcmin. The difference in line style represents the variance expected for galaxies brighter than the specified rest-frame UV magnitudes (\textit{left}) and observed-frame F200W magnitudes (\textit{right}). See text for detailed descriptions of redshift binning and volume normalization across lightcones with different sizes.
	}
	\label{fig:cosmic_variance}
\end{figure*}

As detailed in \citet{Somerville2004}, the relative cosmic variance (with shot noise removed) is defined as
\begin{equation}
	\sigma^2_v \equiv \frac{\langle N^2\rangle - \langle N \rangle^2}{\langle N \rangle^2} - \frac{1}{\langle N \rangle} \text{,}
\end{equation}
where $\langle N \rangle$ and $\langle N^2 \rangle$, which denote the mean and variance of object count $N$, respectively, are the first and second moments of the probability distribution function $P_N(V)$.
Given that the variance is sensitive to volume (and hence also survey area), we adopted the following modified calculation 
\begin{equation}
	\sigma^2_v = \frac{\langle n^2\rangle - \langle n \rangle^2}{\langle n \rangle^2} - \frac{1}{V \langle n \rangle} \text{,}
\end{equation}
where $n$ is the number density of galaxies within a redshift slice and $V$ is the volume. 
In this exercise, we bin galaxies in each lightcone by redshift with bin width $\Delta z = 0.5$ over the range $4 \lesssim z \lesssim 10$.
In this calculation, $V$ is fixed at an assigned value such that $V \langle n \rangle > 1$ even in subregions with extremely low number density (e.g. $n < 1$ when $N > 0$).
In Fig.~\ref{fig:cosmic_variance}, we show the calculated field-to-field variance as a function of redshift for galaxies above several thresholds in rest-frame UV magnitude or observed-frame magnitude in the NIRCam F200W band.
Based on the full set of 40 realizations of wide-field lightcones, We show that the expected fractional root variance evolves from $\sigma_V \sim 0.04$ to $\sim0.27$ from $z\sim4$ to 10 for galaxies with $m_\text{F200W} < 29$ in fields with area $\sim 1000$\,\sqarcmin.

We note that the full set of wide-field lightcones provides 40 example fields for our field-to-field variance calculation, but this would benefit from a larger number of realizations to provide better statistics.
The impact on field-to-field variance from survey area and depth will be further explored in \citet{Yung2022a}, which leverages the robust statistics provided by a set of larger area 2-deg$^2$ lightcones.

\section{Discussion}
\label{sec:discussion}

These simulated lightcones enable a whole new set of applications which will facilitate both planning for future observational programs, and interpreting observational results. In the previous papers in this series, we presented predictions for rest- and observed-frame luminosity functions (\citetalias{Yung2019}); stellar mass functions, stellar-to-halo mass ratios, and other important physical properties (\citetalias{Yung2019a}); ionizing photon production efficiency (\citetalias{Yung2020}) and their subsequent impact on cosmic hydrogen reionization (\citetalias{Yung2020a}); and luminosity functions for active galactic nuclei and their subsequent impact on cosmic helium reionization (\citetalias{Yung2021}). In this work we exploit the lightcones to add predictions of galaxy clustering and field-to-field variance. The lightcones are made publicly available, to enable explorations of the effect of survey size and depth for any desired survey configuration. Furthermore, this enables studies of the evolution of galaxy populations over a continuous redshift distribution, reflecting the evolutionary effects in real surveys with broad redshift bins.

\subsection{Lightcones compared with snapshots}

We first compared one-point distribution functions for rest-frame UV luminosity and observed frame magnitude as predicted by our new lightcones and by our previous approach of utilizing a grid of halos at a fixed output redshift. As expected, the distributions for the rest-UF luminosity agree quite well, although one can see that in some cases there is significant evolution over the redshift range spanned by the bin (e.g. $8 < z < 9$ and $9 < z < 10$ in Fig.~\ref{fig:restUV_LFs}). Larger differences are seen in the distributions of observed frame magnitude, especially in the lowest redshift bin that we show ($3.5 < z < 4.5$), which is due to the effects of k-corrections. This comparison allows estimates of the magnitude of the effect of a continuous redshift distribution on observables compared to using a fixed snapshot in redshift, as is frequently done in comparisons with cosmological hydrodynamic simulations.

\subsection{Angular Clustering at low, medium, and high redshift}

Our main new result is the presentation of measurements of the projected (angular) two-point correlation function (2PCF) both for the CANDELS observations and for our mock lightcones, over a broad redshift range. At low redshift ($0.2 < z < 1.2$), we compare the projected correlation function $\mathit{w}_p(r_p)$ estimate from the PRIMUS and DEEP2 redshifts surveys published by \citet{Skibba2015} with our estimates based on the five observed CANDELS fields for the same stellar mass limit. To our knowledge, the angular correlation functions from CANDELS have not been published before. Our estimate does not attempt to account for the variable depth across the CANDELS fields; however, we select regions for which the depth is reasonably uniform (see S21 for details). Considering the very different areas of these surveys, the different methods used to estimate stellar masses, and the very different precision of the redshift estimates (prism-based or spectroscopic in PRIMUS and DEEP2 vs. predominantly photometric in CANDELS), we found the agreement very encouraging. Unsurprisingly, we find a fairly large field-to-field variance in  $\mathit{w}_p(r_p)$ across the five CANDELS fields, particularly in the smaller volume lower redshift bin. We also compare with the predictions for $\mathit{w}_p(r_p)$ from our mock lightcones, again adopting the same stellar mass limit, but not attempting to mimic specific survey characteristics or to include observational effects such as detailed selection or redshift uncertainties. Again, we find remarkably good agreement between the estimates based on our mock lightcones and the observational estimates, mostly within the uncertainties of the observed CANDELS fields and within a factor of 2 in the small-scale end.

In addition, we compared the angular correlation function (ACF) that we measured from the observed CANDELS catalogues (again, from all five fields) with that measured from our mock lightcones for intermediate redshifts ($1.25 < z < 4.5$, in four redshift bins). Again, considering the relative crudeness of the comparison, we were encouraged by the agreement. Referring to Fig.~\ref{fig:mid-z_clustering}, we clearly see the change in slope of the ACF at around 4-10 arcsecond which is thought to be due to the transition from domination by the `one halo' term (galaxies that reside within the same dark matter halo) to the `two-halo' term (galaxies residing in separate host halos). That this transition occurs at roughly the same scale in both the observational and theoretical estimates is particularly encouraging (the contribution of the one- and two-halo term to our predicted ACF will be quantified in more detail in \citet{Yung2022a}). We note in this context that the modelling of satellite galaxies in SAMs (which produce the one-halo term) is particularly uncertain (see Section~\ref{sec:caveats} for a discussion).

Lastly we compare the estimates of the angular correlation function at high redshift ($4 < z < 7$) from our mock lightcones with observational estimates from HST and Subaru \citep{Harikane2016}. Once again, considering the many uncertainties, predictions from our lightcones are well within the errors provided for these observational constraints, including a hint of the transition from the one- to two-halo term.

We emphasize that these are genuine predictions of our model, as no information about clustering is used to calibrate the model. It has been demonstrated before that when galaxy global properties (stellar mass or luminosity) are mapped to halo mass using abundance matching, good agreement with observed clustering is obtained \citep{Conroy2009a, Behroozi2013a, Behroozi2019, Behroozi2020, Moster2013, Moster2018}. These previous studies have generally focussed on low to intermediate redshift.  We have shown previously that our model predictions for luminosity functions and stellar mass functions are in good agreement with existing observations up to $z\sim 8$--9 \citep{Yung2019,Yung2019a}. However, the good agreement of our model predictions with observed clustering measurements is an independent indication that the mapping between halo mass and galaxy stellar mass or luminosity assigned by the physical prescriptions in our models is consistent with that in the real Universe. Given the large uncertainties in dust corrections and stellar mass estimates at high redshift, this is an important confirmation.

Encouraged by these results, we present forecasts for the angular correlation function measurements that may be obtained with future \emph{JWST} surveys with a range of areas and depths. Additionally, using the many realizations of our mock lightcones, we illustrate the field-to-field variance in measurements from surveys with an area of $\sim 1000$ arcmin$^2$. Our forecasts suggest that we may eventually be able to obtain robust measurements of angular clustering out to redshifts of 8 or 9 with \emph{JWST}.

In principle, one can attempt to estimate the host halo mass of a galaxy population based on its observed clustering, using a halo occupation distribution (HOD) formalism \citep{Berlind2002, Zehavi2004, Zehavi2011, Zheng2007, Harikane2016}.
This can then be used as an alternative method for constraining the relationship between galaxy luminosity or stellar mass and dark matter halo mass. Our mocks can be used to forecast the expected accuracy of such estimates based on specific survey characteristics. We plan to explore this in a future work.

\subsection{A tool for detailed estimates of field-to-field variance}

The multiple realizations of lightcones with a range of area and depth provide a powerful tool to the community for estimating field-to-field variance in a side variety of observable quantities in currently planned or future surveys. We have deliberately provided lightcones with larger areas than anticipated near-term \emph{JWST} survey projects, so that they can be cut to any desired geometry. We emphasize that the quantitative results for field-to-field variance presented here do \emph{not} correspond to specific observational surveys, but represent the expected variance over our suite of mock lightcones.
Field-to-field variance can also be estimated using analytic prescriptions \citep{Somerville2004, Trenti2008, Moster2011} and in hydrodynamic simulations \citep{Bhowmick2020}. \citet{Endsley2020} conducted a related study based on lightcones filled with galaxies from the empirical \textsc{UniverseMachine} model and explored the impact on clustering predictions due to sample variance.
Predictions for how field-to-field variance for the ACF and SFRD vary with survey area will be explored in more detail in \citet{Yung2022a}.

\subsection{Caveats and limitations of current models}
\label{sec:caveats}
The limitations and caveats regarding the baryonic processes in the SAM and in galaxies at $z \gtrsim 10$ have been thoroughly discussed in previous works; we refer the reader to section 6.3 in \citetalias{Yung2019a} and section 4.2 in \citetalias{Yung2020a}.
This discussion will be focused mainly on the topics related to the construction of simulated lightcones and clustering predictions presented in this work.

We utilize Extended Press-Schechter based merger trees to populate halos in our lightcones with galaxies. The EPS-based merger trees adopted in this work series have been compared with trees extracted from numerical simulations and the results shown to be in good agreement. However, the EPS algorithm has never been tested over the full halo mass and redshift ranges that are explored in this work, as there is currently no publicly available relevant suite of dark matter only simulations. Furthermore, the EPS-based merger tree approach reconstructs only an approximate reconstruction of the joint distribution of halo progenitor masses \citep{Somerville2000}.

Moreover, it is known that in N-body simulations, the merger histories of halos depend on their large scale environmental density \citep{AvilaReese2005, Maulbetsch2007, Wang2007, Fakhouri2009, Fakhouri2010}. This second-order dependence is not captured by any of the current semi-analytic merger tree building algorithms. Because galaxy properties such as stellar mass are known to depend strongly on halo formation history in SAMs \citep{Gabrielpillai2021}, this implies that the mapping between stellar mass and halo mass predicted by the SAMs presented here will not properly reflect this secondary dependence on large scale environment, which has implications for the predicted clustering. Similarly, each halo is processed independently with no `knowledge' of its larger scale environment. Other processes, such as metal enrichment or radiation fields could introduce an environmental dependence on galaxy physical properties, which will not be captured by our current models.

The properties of satellite galaxies are particularly uncertain in our SAM framework. The SC SAM includes a semi-analytic model that describes the decay of satellite orbits due to dynamical friction, and the tidal stripping and destruction of these objects as they orbit within the halo. However, this modelling is highly uncertain and does not include the effects of stripping and destruction due to strong encounters with the baryonic components of galaxies, which can have a large effect on tidal stripping and destruction \citep{Dooley2016, Smith2016, Han2018}. The satellite population is also particularly sensitive to inaccuracies in the EPS-based halo merger trees discussed above. Furthermore, we currently assume that when a galaxy becomes a satellite, its hot gas reservoir is instantaneously stripped off, robbing the galaxy of a supply of new fuel for star formation. The satellite galaxy continues to eject gas from its ISM via stellar feedback, but this gas is assumed to be added to the hot gas reservoir of the central galaxy. These assumptions may lead to satellite galaxies being prematurely starved of fresh gas and quenching too early \citep[e.g.][]{Kimm2009}. Finally, we have found that the prescription for tidal stripping and destruction in our models does not yield a radial distribution of satellites that is in agreement with the predictions from N-body simulations, so we have reassigned satellite positions in post-processing, making the simple assumption that the radial profile of satellites traces the overall dark matter density profile as represented by an NFW profiles. This is known not to be the case in detail in hydrodynamic simulations \citep{Bose2020, Bose2022, McDonough2022}.

Current photometry are computed based on composite stellar spectra from \citetalias{Bruzual2003}. The photometry in the released mock lightcones include contributions from the stellar continuum only, and do not include the emission from nebular lines or continuum, nor radiation from AGN. In most cases, these contributions will be negligible. However, in a future work, we will implement emission line predictions from models presented by \citet{Hirschmann2017, Hirschmann2019} and explore their effects on broad-, medium-, and narrow band photometry (Yung, Hirschmann, Somerville et al., in prep). A larger source of uncertainty is the impact of attenuation from ISM dust. We have adopted a simple, empirical approach in which we adjust the normalization of the optical depth at the $V$ band in order to match the observed UV luminosity function, and we adopt a fixed functional form for the attenuation as a function of wavelength (fixed attenuation curve). The finding that we need to adjust the `fudge factor' for the dust optical depth as an ad hoc function of wavelength is a good indication that we do not yet have a good physical handle on the evolution of dust properties. We note that many other studies have similarly found that an ad hoc adjustment of the dust optical depth normalization is needed to match observations \citep[e.g.][]{Vogelsberger2020a}.
We also plan on improving this component of our modelling in future work. Lastly, we have adopted the \citet{Madau1996} prescription for attenuation by the IGM throughout our lightcones, while it is only applicable for a uniform IGM, i.e. post-reionization. Improving on this would require full radiative transfer.

\subsection{The role of \emph{JWST forecasts} in survey planning}

These lightcones have been used extensively in developing upcoming \emph{JWST} surveys CEERS and NGDEEP surveys, where the simulated datasets are used to inform survey specifications, including number of objects expected for a given survey area and exposure time. The simulated catalogues are also further processed into mock images, which are useful for practising data reduction, source extraction, removal of dithering pattern, and optimizing colour selection criteria. At the time when \emph{JWST} observations become available, the theory framework presented in the work series will be utilized for the physical interpretation of detected sources.

While \emph{JWST} will be the powerhouse for high-redshift observations, its relatively narrow field of view restricts its ability to explore large-scale structures. Selected \emph{JWST} Cycle 1 program COSMOS-Web \citep{Kartaltepe2021} is expected to survey $\sim 0.6$\,\sqdeg\ of the sky, which is expected to be larger than any existing contiguous fields surveyed by \emph{Hubble} while reaching a survey depth comparable to \emph{HUDF}. 
Current generation of \emph{JWST} wide-field surveys will set up for future deep-field with \emph{Roman Space Telescope}, which IR sensitivity is comparable to \emph{Hubble} and a single pointing of its on-board Wide-Field Instrument (WFI) will span $\sim0.28$\,\sqdeg.
\emph{JWST} observations will also shape future high-redshift surveys with ESA's \emph{Euclid} Observatory. A companion paper \emph{Semi-analytic forecasts for Roman} \citep{Yung2022a} will present a set of 2-deg$^2$ lightcones that are constructed with the same infrastructure that produces the lightcones presented in this work and will support future wide-field surveys and the potential synergy across multiple instruments.

\section{Summary and conclusions}
\label{sec:snc}

In this work, we present mock galaxy catalogues that are optimized to support the planning and interpretation of \emph{JWST} surveys. 
These mock catalogues come in two flavours: wide-field and deep-field.
The set of wide-field lightcones consists of five independent fields with coordinates and geometries overlapping the five CANDELS legacy fields, spanning $z\sim0$ to 10 and resolving down to observed-frame $m_\text{F200W}\sim 30$ or rest-frame $M_\text{UV}\sim -18$.
The set of ultra-deep lightcones represents a 132\sqarcmin\ field overlapping HUDF, spanning range $z\sim0$ to $\sim12$ and resolving down to observed-frame $m_\text{F200W}\sim 34$ or rest-frame $M_\text{UV}\sim -14$.
Eight realizations have been created for each field by sampling different halos from hydrodynamic simulations, which propagates physically accurate cosmic variance to the simulated fields.
A total of 48 lightcones presented in this work provides a total of $\simeq12.2$\sqdeg\ of (non-contiguous) simulated area.

The construction of these past lightcones leverages both the accurate representation of structure formation in cosmological N-body simulations and the efficiency of semi-analytic models.
Dark matter halos sourced from state-of-the-art cosmological simulations are projected onto past lightcones that are traced by the lines of sight within the area of a specified field.
The merger histories of individual halos are constructed on-the-fly using the EPS formalism, within which the formation and evolution of galaxies and a wide range of associated physical processes are predicted by the well-established Santa Cruz SAM.
In previous papers in the \textit{Semi-analytic Forecasts} series, the SAM has been shown to reproduce a variety of observational constraints at high redshift, including UV luminosity functions, stellar mass functions, star formation rate functions, and stellar-to-halo mass ratios. 
We also present predictions that are specifically tailored to \emph{JWST}, including galaxies in the luminosity (or mass) range that have not been detected before, as well as NIRCam broad- and medium-band photometry for all galaxies.

In this final paper of the \textit{Forecasts} series, we take full advantage of the mock lightcones and present new predictions that take advantage of the lightcone format, including continuous redshift evolution of galaxies and their spatial distributions. 
We examine the clustering of galaxies in these mock lightcones by computing projected two-point correlation functions and compare them to existing measurements from past galaxy surveys.
Our results are in excellent agreement with past observations across a wide range of redshifts.
Utilizing the two sets of simulated lightcones covering areas and depths that are similar to future \emph{JWST} wide and deep surveys, we provide \emph{forecasts} for two-point correlation functions that may be obtained from future \emph{JWST} surveys.

We summarize our main conclusions below.
\begin{enumerate}
	\item We assembled a modelling pipeline that sources halos from N-body simulations along a past lightcone and populates them with galaxies with a wide range of predicted properties. The set of mock wide-field and ultra-deep lightcones delivers a wide variety of predictions for upcoming \emph{JWST} surveys, including one-point distribution functions of physical and observable prosperities, as well as time-evolution of galaxy populations across different epochs and clustering statistics with two-point auto-correlation functions.
	
	\item The wide-field lightcones cover $\sim 1000$\sqarcmin\ each, containing galaxies with $-16 > M_\text{UV} > -22$ between $0 \lesssim z \lesssim 10$. The ultra-deep lightcones cover $132$\,\sqarcmin\ each, containing galaxies $-12 > M_\text{UV} > -21$ between $0 \lesssim z \lesssim 12$.
	
    \item The predicted clustering of galaxies in our simulated lightcones is in excellent agreement with observed clustering over a wide redshift range $0.2 \lesssim z \lesssim 7.5$.
    
    \item We show that \emph{JWST} wide-field surveys of $\sim 1000$\,\sqarcmin\ with depth reaching $m_{F200W}\sim29$ will be capable of constraining galaxy clustering up to an angular separation of $\sim 10$ arcsecond.

	\item We show that the expected field-to-field variance in the angular correlation function at separation scale $\sim100$ arcseconds can be up to a factor of 2 for galaxies with $m_\text{F200W} < 29$ in fields with area $\sim 1000$\,\sqarcmin.

    \item We quantify the expected field-to-field variance in the cosmic star formation rate density (SFRD) for galaxies with $M_\text{UV} < -17$ in fields with area $\sim 1000$\,\sqarcmin\ to be $\sim 10\%$ at $z\sim4$ and up to $\sim 35\%$ at $z\sim10$.
    
    \item Based on the full set of 40 realizations of wide-field lightcones, we show that the expected fractional root variance evolves from $\sigma_V \sim 0.04$ to $\sim0.27$ from $z\sim4$ to 10 for galaxies with $m_\text{F200W} < 29$ in fields with area $\sim 1000$\,\sqarcmin.
\end{enumerate}

The results presented in this paper are intended as just a few examples of the predictions that can be extracted from these lightcones. All of the lightcones are made publicly available so that the community can exploit them for numerous additional applications.

\section*{Acknowledgements}

The authors of this paper would like to thank Austen Gabrielpillai, Yuichi Harikane, Shengqi Yang, and Ulrich Steinwandel for useful discussions that helped shape this work. 
We also thank the members of CEERS and NGDEEP teams for utilizing the pre-production results and providing feedback that improved this work. 
We thank the anonymous referee for the constructive comments that improved this work.
The simulations and data products for this work were run on computing machines, \textit{astera} and \textit{seliana}, managed by the Office of Scientific Computing at NASA Goddard Space Flight Center.
We thank the Center for Computational Astrophysics (CCA) and the Scientific Computing Core (SCC) at the Flatiron Institute for hosting the data associated with this work on the data release portal \textit{Flathub}. 
We warmly thank Dylan Simon and Elizabeth Lovero for coordinating the data release portal and project website.
AY is supported by an appointment to the NASA Postdoctoral Program (NPP) at NASA Goddard Space Flight Center, administered by the Oak Ridge Associated Universities under contract with NASA. 
AY also thanks the Center for Computational Astrophysics at the Flatiron Institute for hospitality during this work.
RSS acknowledges support from the Simons Foundation.

Use of the mock catalogues presented in this work should reference both this work and \citealt{Somerville2021} for the construction of the simulated lightcones, and reference \citealt{Somerville2015} for the Santa Cruz semi-analytic model and \citealt{Yung2019} for model calibration and validation against existing observation constraints.

\section*{Data Availability}
The data underlying this paper are available in the Data Product Portal hosted by the Flatiron Institute at\\ \url{https://www.simonsfoundation.org/ semi-analytic-forecasts-for-jwst/} and the Flatiron Institute Data Exploration and Comparison Hub (Flathub, \url{http://flathub.flatironinstitute.org/group/sam-forecasts}).



\bibliographystyle{mnras}
\bibliography{library.bib} 



\appendix

\section{Tabulated values for selected two-point correlation functions}
\label{appendix:a}
\setcounter{table}{0} \renewcommand{\thetable}{A\arabic{table}}
We provide tabulated data for selected angular two-point correlation functions with limiting magnitude $m_\text{F200W,lim}\sim29$ and $\sim31$ in Tables \ref{table:angular} and \ref{table:angular2}, respectively.

\begin{table*}
	\centering
	\caption{This tables provide the tabulated angular correlation function computed from the wide-field lightcones for galaxies $m_\text{F200W,lim}\sim29$. We provide the median value and the 16th and 84th percentile across the 40 wide-field lightcones.}
	\label{table:angular}
	\begin{tabular}{ccccccc}
		\hline
		& \multicolumn{6}{c}{$\log\,w(\theta)$}\\
		$\log(\theta$/arcsec)  & $z = 3.50$ -- 4.50 & $z = 4.50$ -- 5.50 & $z = 5.50$ -- 6.50 & $z = 6.50$ -- 7.50 & $z = 8.00$ -- 9.00 & $z = 9.00$ -- 10.00    \\
		\hline
		0.47 & $-0.472^{+0.036}_{-0.028}$ & $-0.243^{+0.036}_{-0.039}$ & $0.014^{+0.042}_{-0.054}$ & $0.307^{+0.052}_{-0.056}$ & $0.814^{+0.152}_{-0.201}$ &   \\[0.07in]
		0.67 & $-0.569^{+0.028}_{-0.036}$ & $-0.371^{+0.035}_{-0.025}$ & $-0.151^{+0.045}_{-0.048}$ & $0.09^{+0.034}_{-0.03}$ & $0.562^{+0.113}_{-0.075}$ &   \\[0.07in]
		0.87 & $-0.739^{+0.027}_{-0.036}$ & $-0.559^{+0.048}_{-0.037}$ & $-0.344^{+0.045}_{-0.039}$ & $-0.089^{+0.045}_{-0.058}$ & $0.327^{+0.132}_{-0.105}$ & $0.684^{+0.219}_{-0.109}$ \\[0.07in]
		1.07 & $-0.918^{+0.036}_{-0.036}$ & $-0.716^{+0.047}_{-0.049}$ & $-0.507^{+0.045}_{-0.059}$ & $-0.291^{+0.068}_{-0.054}$ & $0.107^{+0.162}_{-0.113}$ & $0.452^{+0.206}_{-0.089}$ \\[0.07in]
		1.27 & $-1.089^{+0.071}_{-0.023}$ & $-0.858^{+0.057}_{-0.056}$ & $-0.652^{+0.036}_{-0.064}$ & $-0.467^{+0.062}_{-0.042}$ & $-0.072^{+0.096}_{-0.121}$ & $0.201^{+0.151}_{-0.11}$ \\[0.07in]
		1.47 & $-1.23^{+0.096}_{-0.035}$ & $-0.993^{+0.066}_{-0.086}$ & $-0.815^{+0.054}_{-0.053}$ & $-0.617^{+0.08}_{-0.099}$ & $-0.31^{+0.207}_{-0.102}$ & $-0.008^{+0.115}_{-0.162}$ \\[0.07in]
		1.67 & $-1.346^{+0.107}_{-0.061}$ & $-1.121^{+0.099}_{-0.123}$ & $-0.955^{+0.076}_{-0.069}$ & $-0.789^{+0.111}_{-0.105}$ & $-0.488^{+0.186}_{-0.145}$ & $-0.277^{+0.153}_{-0.271}$ \\[0.07in]
		1.87 & $-1.482^{+0.149}_{-0.109}$ & $-1.263^{+0.121}_{-0.131}$ & $-1.1^{+0.06}_{-0.107}$ & $-0.972^{+0.149}_{-0.143}$ & $-0.71^{+0.229}_{-0.31}$ & $-0.557^{+0.23}_{-0.404}$ \\[0.07in]
		2.07 & $-1.643^{+0.19}_{-0.123}$ & $-1.402^{+0.143}_{-0.179}$ & $-1.293^{+0.09}_{-0.109}$ & $-1.187^{+0.205}_{-0.176}$ & $-0.876^{+0.231}_{-0.388}$ & $-0.756^{+0.173}_{-0.371}$ \\[0.07in]
		2.27 & $-1.825^{+0.179}_{-0.142}$ & $-1.552^{+0.227}_{-0.306}$ & $-1.527^{+0.14}_{-0.173}$ & $-1.452^{+0.297}_{-0.271}$ & $-1.152^{+0.431}_{-0.691}$ & $-0.969^{+0.206}_{-1.526}$ \\[0.07in]
		2.47 & $-2.043^{+0.272}_{-0.504}$ & $-1.733^{+0.251}_{-0.493}$ & $-1.804^{+0.219}_{-0.328}$ & $-1.77^{+0.459}_{-0.669}$ & $-1.453^{+0.457}_{-0.799}$ &   \\[0.02in]
		\hline
	\end{tabular}
\end{table*}

\begin{table*}
	\centering
	\caption{This tables provide the tabulated angular correlation function computed from the ultra-deep lightcones for galaxies $m_\text{F200W,lim}\sim31$. We provide the median value and the 16th and 84th percentile across the set of eight lightcone realizations.}
	\label{table:angular2}
	\begin{tabular}{ccccccccc}
		\hline
		& \multicolumn{6}{c}{$\log\,w(\theta)$}\\
		$\log(\theta$/arcsec)  & $z = 3.50$ -- 4.50 & $z = 4.50$ -- 5.50 & $z = 5.50$ -- 6.50 & $z = 6.50$ -- 7.50 & $z = 8.00$ -- 9.00 & $z = 9.00$ -- 10.00    \\
		\hline
		0.27 & $-0.326^{+0.043}_{-0.021}$ & $-0.377^{+0.012}_{-0.036}$ & $-0.468^{+0.117}_{-0.112}$ & $-0.324^{+0.041}_{-0.04}$ & $0.104^{+0.232}_{-0.072}$ &   \\[0.07in]
		0.47 & $-0.716^{+0.016}_{-0.017}$ & $-0.728^{+0.06}_{-0.054}$ & $-0.59^{+0.111}_{-0.147}$ & $-0.304^{+0.08}_{-0.034}$ & $0.097^{+0.042}_{-0.063}$ & $0.438^{+0.146}_{-0.292}$ \\[0.07in]
		0.67 & $-0.986^{+0.026}_{-0.083}$ & $-0.849^{+0.047}_{-0.064}$ & $-0.569^{+0.046}_{-0.091}$ & $-0.375^{+0.032}_{-0.095}$ & $-0.009^{+0.106}_{-0.095}$ & $0.279^{+0.154}_{-0.162}$ \\[0.07in]
		0.87 & $-1.156^{+0.044}_{-0.041}$ & $-0.915^{+0.074}_{-0.07}$ & $-0.723^{+0.109}_{-0.128}$ & $-0.551^{+0.092}_{-0.036}$ & $-0.209^{+0.062}_{-0.107}$ & $-0.057^{+0.176}_{-0.036}$ \\[0.07in]
		1.07 & $-1.254^{+0.031}_{-0.059}$ & $-1.051^{+0.105}_{-0.123}$ & $-0.904^{+0.13}_{-0.035}$ & $-0.67^{+0.019}_{-0.156}$ & $-0.383^{+0.08}_{-0.124}$ & $-0.212^{+0.123}_{-0.183}$ \\[0.07in]
		1.27 & $-1.413^{+0.075}_{-0.039}$ & $-1.182^{+0.117}_{-0.182}$ & $-1.041^{+0.123}_{-0.081}$ & $-0.832^{+0.042}_{-0.102}$ & $-0.545^{+0.095}_{-0.164}$ & $-0.427^{+0.096}_{-0.096}$ \\[0.07in]
		1.47 & $-1.527^{+0.087}_{-0.037}$ & $-1.309^{+0.077}_{-0.195}$ & $-1.168^{+0.097}_{-0.119}$ & $-0.968^{+0.046}_{-0.243}$ & $-0.752^{+0.101}_{-0.106}$ & $-0.67^{+0.106}_{-0.08}$ \\[0.07in]
		1.67 & $-1.654^{+0.059}_{-0.077}$ & $-1.489^{+0.118}_{-0.197}$ & $-1.357^{+0.19}_{-0.021}$ & $-1.138^{+0.108}_{-0.268}$ & $-0.973^{+0.163}_{-0.122}$ & $-0.884^{+0.097}_{-0.142}$ \\[0.07in]
		1.87 & $-1.847^{+0.155}_{-0.085}$ & $-1.782^{+0.321}_{-0.245}$ & $-1.491^{+0.174}_{-0.345}$ & $-1.271^{+0.044}_{-0.627}$ & $-1.35^{+0.252}_{-0.216}$ & $-1.45^{+0.4}_{-0.28}$ \\[0.07in]
		2.07 & $-2.148^{+0.354}_{-0.173}$ & $-1.93^{+0.31}_{-1.384}$ & $-1.763^{+0.374}_{-0.381}$ & $-1.582^{+0.082}_{-0.379}$ &   &   \\[0.02in]
		\hline
	\end{tabular}
\end{table*}

\section{Tabulated values for cosmic SFR}
\label{appendix:b}
\setcounter{table}{0} \renewcommand{\thetable}{B\arabic{table}}
We provide tabulated data for the predicted cosmic SFR presented in Fig.~\ref{fig:CSFR_all_fields}. We provide values from the median (red solid line) and the 84th and 16th percentile (red dashed lines).

\begin{table*}
	\centering
	\caption{This tables provide the tabulated cosmic SFR predicted by the Santa Cruz SAM. We provide the median value and 16th and 84th percentile across the 40 wide-field lightcones (see Fig.~\ref{fig:CSFR_all_fields}) between $z = 2$ to 10 and the same quantity calculated from a grid run between $z = 11$ to 15 from \citetalias{Yung2020a}. These predictions are made with the same fiducial model.}
	\label{table:csfr}
	\begin{tabular}{ccccccc}
		\hline
           	   & This work                                 &        & This work                                 &        & \citetalias{Yung2020a}                    \\
	    $z$    & $\log$(CSFR/\Msun\,yr$^{-1}$\,Mpc$^{-3}$) &  $z$   & $\log$(CSFR/\Msun\,yr$^{-1}$\,Mpc$^{-3}$) &  $z$   & $\log$(CSFR/\Msun\,yr$^{-1}$\,Mpc$^{-3}$) \\
		\hline\\[-0.12in]
		2.0 -- 2.5 & $-1.01^{+0.05}_{-0.05}$ & 6.0 -- 6.5 & $-1.69^{+0.06}_{-0.11}$ & 11 & $-3.94$ \\[0.07in]
		2.5 -- 3.0 & $-0.98^{+0.03}_{-0.04}$ & 6.5 -- 7.0 & $-1.86^{+0.06}_{-0.02}$ & 12 & $-4.63$ \\[0.07in]
		3.0 -- 3.5 & $-0.99^{+0.07}_{-0.05}$ & 7.0 -- 7.5 & $-2.09^{+0.05}_{-0.06}$ & 13 & $-5.31$ \\[0.07in]
		3.5 -- 4.0 & $-0.99^{+0.05}_{-0.04}$ & 7.5 -- 8.0 & $-2.26^{+0.07}_{-0.15}$ & 14 & $-6.49$ \\[0.07in]
		4.0 -- 4.5 & $-1.09^{+0.04}_{-0.05}$ & 8.0 -- 8.5 & $-2.51^{+0.05}_{-0.05}$ & 15 & $-7.73$ \\[0.07in]
		4.5 -- 5.0 & $-1.25^{+0.06}_{-0.05}$ & 8.5 -- 9.0 & $-2.74^{+0.08}_{-0.10}$ \\[0.07in]
		5.0 -- 5.5 & $-1.36^{+0.05}_{-0.04}$ & 9.0 -- 9.5 & $-2.98^{+0.12}_{-0.05}$ \\[0.07in]
		5.5 -- 6.0 & $-1.54^{+0.07}_{-0.09}$ & 9.5 -- 10.0 & $-3.26^{+0.11}_{-0.19}$ \\[0.02in]
		\hline
	\end{tabular}
\end{table*}

\bsp	
\label{lastpage}
\end{document}